\documentclass[twocolumn,aps,rmp]{revtex4}
\usepackage{amsfonts,amssymb,amsmath,bm}
\usepackage[dvips]{graphicx}
\usepackage{url}
\usepackage{wasysym}
\newcommand{\1}{\mathrm{d}}
\newcommand\ds{\displaystyle}
\newcommand\ttaurus{\pmb{\taurus}}
\begin{document}

\title{TRANSLATION OF LEONHARD EULER'S: GENERAL PRINCIPLES OF THE MOTION OF
  FLUIDS}%
\author{U. Frisch}
\affiliation{Labor. Cassiop\'ee, UNSA, CNRS, OCA, BP 4229,
06304 Nice Cedex 4, France}

\begin{abstract}
This is an adapatation by U.~Frisch of an English translation
  by Thomas~E.~Burton of Euler's memoir `Principes g\'en\'eraux du
  mouvement des fluides' (Euler, 1775b).  Burton's translation
  appeared in \textit{Fluid Dynamics} {\bf 34} (1999) pp.~801--82,
  Springer and is here adapted by permission.  A detailed presentation
  of Euler's published work can be found in Truesdell, 1954. Euler's
  work is discussed also in the perspective of eighteenth century
  fluid dynamics research by Darrigol and
  Frisch, 2008.
  Explanatory footnotes have been supplied where necessary by
  G.K.~Mikhailov and a few more by U.~Frisch and O.~Darrigol. Euler's memoir had neither
  footnotes nor a list of references.
\end{abstract}
\maketitle

 {\bf 1}.~~~Having established in my previous Memoir\footnote{Euler, 1755a} the
 principles of fluid equilibrium in their most general form, regarding
 both the diverse nature of fluids and the forces that act upon them,
 I now propose to deal with the motion of fluids in the same way and
 to seek out the general principles on which the entire science of
 fluid motion is based.  It will readily be understood that this is a
 much more difficult undertaking and involves studies of incomparably
 greater depth. Nevertheless, I hope to arrive at an equally
 successful conclusion, so that, if difficulties remain, they will
 pertain not to Mechanics but purely to Analysis, this science not yet
 having been brought to the degree of perfection necessary to develop
 analytical equations [\textit{formules}]\footnote{Bracketed words are from the
 original eighteenth century French text.} that embody the principles
 of fluid motion.\\

 {\bf 2}.~~~The task, then, is to discover the principles by means of which the
 motion of a fluid can be determined, whatever its state and whatever
 the forces to which it is subjected. To this end, we shall examine
 in detail all the elements which form the subject of our research and
 contain quantities both known and unknown.  First of all, the nature
 of the fluid is assumed to be known, in which case it is necessary to
 consider its various forms since it may be compressible or
 incompressible. If it is not compressible, then there are two
 possibilities: either the entire mass is composed of homogeneous
 parts, whose density is everywhere and always the same, or it is 
composed of heterogeneous parts and in this case it is necessary to know
 the density of each component and the proportions of the mixture. If
 the fluid is compressible and its density is variable, we must know
 the law according to which its elasticity\footnote{By elasticity
 [\textit{\'elasticit\'e}] Euler means that property of a fluid which is
 expressed in the creation of 
internal pressure and therefore uses 
the term on an equal footing with the term ``pressure'' (see \S~5
 below).} 
depends on the density
 and whether the elasticity depends only on the density or also on
 some other property, such as heat,\footnote{Essentially, heat [\textit{chaleur}] should be taken to mean temperature.} which is proper to each particle
 of fluid, at least for each instant of time.\\

 {\bf 3}.~~~It must also be assumed that the state of the fluid at a
certain moment of time is known and I shall call this the initial
state [\textit{\'etat primitif}] of the fluid. As this state is quasi-arbitrary,
it is necessary, first of all, to know the distribution of the
particles of which the fluid is composed and, unless in the initial
state the fluid is at rest, the motion impressed upon them. However,
the initial motion is not entirely arbitrary since both the continuity
and the impenetrability of the fluid impose a certain limitation which
I shall investigate below. Often, however, nothing is known of the
initial state, for example when it is a question of determining the
motion of a river, and then it is usually only possible to seek the
steady state at which the fluid finally arrives, thereafter undergoing
no further changes. Now, neither this circumstance nor the initial
state in any way affect the investigation to be made and the
calculations will always be the same. It is only in the integrations
that they need to be taken into account for the purpose of determining
the constants which every integration involves.
\\

 {\bf 4}.~~~Thirdly, the data must include the external forces to which the
fluid is subjected. I shall call these forces external to distinguish
them from the internal forces which the fluid particles exert on each
other and which will constitute the main topic for subsequent
investigation. Thus, it could be assumed that the fluid is not
exposed to any external force, unless it be natural gravity which is
everywhere considered to be constant in magnitude and to act in the
same direction.  However, to generalize the investigation, I shall
consider the fluid to be acted upon by forces which may be directed
towards one or more centers or obey some other law with respect to
both magnitude and direction. As far as these forces are concerned,
only their accelerating action is directly known, irrespective of the
masses upon which they act.  Accordingly, I shall introduce into the
calculations only the accelerative forces, from which it will be easy
to obtain the true motive forces by multiplying in each case the
accelerative forces by the masses to which they are
applied.\footnote{Newton \label{fn:newton} distinguishes between the ``accelerative'' and
``motive'' aspects of a force, the former being ``a measure proportional
to the velocity which it generates'' and the latter ``a measure
proportional to the quantity of motion which it generates in a given
time''. Thus, the ``accelerative force'' is the ratio of the acting force
to the mass of the particle on which it acts, i.e. the acceleration
which it imparts, and the ``motive force'' is that which, strictly
speaking, we now understand by force. The neutral term ``acting
forces'' [\textit{forces sollicitantes}], not used by Newton, was widely
employed by Euler, starting with his well-known ``Mechanics'' (Euler,
1736).}\\

{\bf 5}.~~~Let us now turn to those elements which contain that which is
unknown. In order properly to understand the motion that will be
imparted to the fluid it is necessary to determine, for each instant
and for each point, both the motion and the pressure [\textit{pression}] of
the fluid situated there. And if the fluid is compressible, it is
also necessary to determine the density, knowing the above-mentioned
other property which, together with the density, makes it possible to
determine the elasticity. The latter, being counterbalanced by the 
fluid pressure, must be considered equal to that pressure, exactly as
in the case of equilibrium, where I have developed these ideas more
thoroughly.\footnote{Cf. Euler, 1755a.} Clearly, then, the number of quantities which enter into
the study of fluid motion is much greater than in the case of
equilibrium, since it is necessary to introduce letters which denote
the motion of each particle and all these quantities may vary with
time. Thus, in addition to the letters which determine the location of
each conceivable point in the fluid, another is required which
denotes the time already elapsed and which, by virtue of its
variability, 
can be applied to any given time.\\

{\bf 6}.~~~Suppose (Fig.~1) that from the initial state a time $t$ has
elapsed and that the fluid is now in a state of motion which is to be
determined.\footnote{In the original publication all figures are on  the
  fourth table following the end (on p.~402) of the part of the volume
  dedicated to 
the Mathematics Class. As was the rule at the time figures are devoid of 
captions.
}  Whatever the volume that the fluid now occupies, I begin
by considering any point Z in the fluid mass and in order to
introduce the location of this point Z into the calculations I relate
it to three fixed axes, OA, OB and OC, mutually perpendicular at the
point O and having a given position. Let the two axes OA and OB lie in
the plane represented by the page and let the third OC be
perpendicular to it. Then from the point Z we draw a perpendicular ZY
to the plane AOB and from the point Y a normal YX to the axis OA to
obtain three coordinates: ${\rm OX}=x$, ${\rm XY}=y$ and ${\rm YZ}=z$
parallel to our three axes. For each point in the fluid mass, these
three coordinates $x$, $y$ and $z$ will have specific values and by
successively giving these three coordinates all possible values, both
positive and negative, we can run through all the points of infinite
space, including those lying in the volume occupied by the fluid at
each instant of time.
\begin{figure}[!h]
  \includegraphics[scale=0.5,angle=-1]{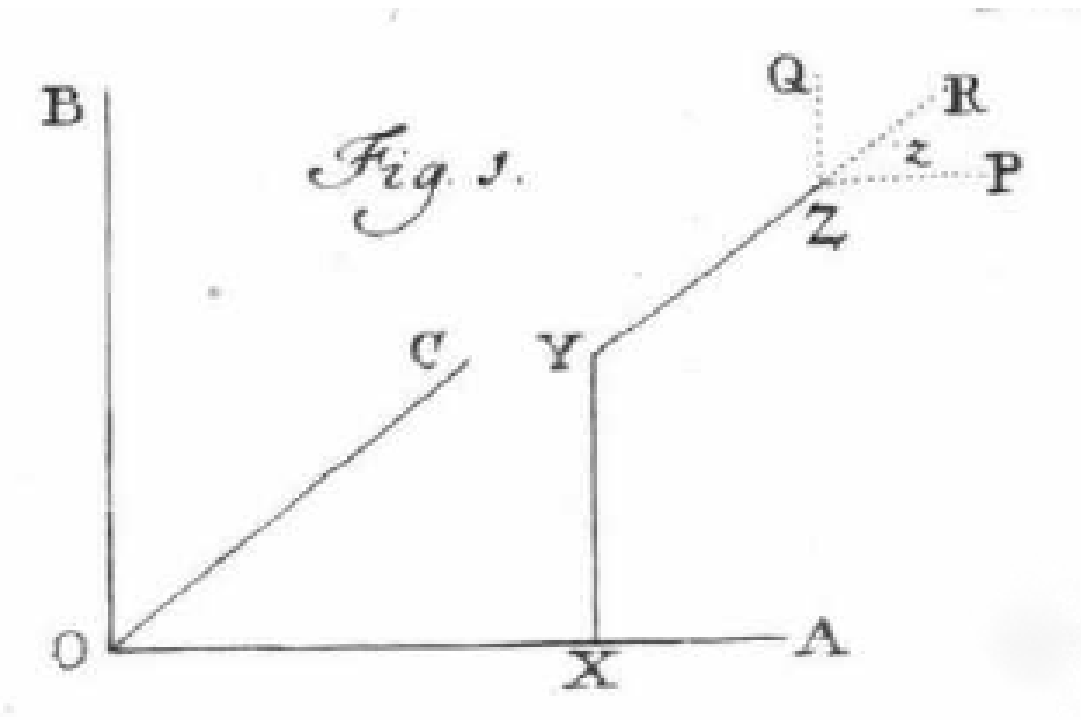}%
\end{figure}
\\

{\bf 7}.~~~Secondly, I shall consider the accelerative forces which act
at a given moment on the fluid particle located at Z.  Now, whatever
these forces may be, they can always be reduced to three acting in the
three directions ZP, ZQ and ZR parallel to our three axes 0A, OB and
OC. Taking the accelerative force of natural gravity\footnote{The acceleration of gravity is intended.} as the unit, we
let P, Q and R be the accelerative forces acting on the point Z in the
directions ZP, ZQ and ZR, the letters P, Q and R denoting abstract
numbers [\textit{nombres absolus}].\footnote{The non-dimensionality of the values of P, Q and R is emphasized.} If unchanging forces always act at the
same point in space Z, the quantities P, Q and R will be expressed by
certain functions of the three coordinates $x$, $y$ and $z$, However, if the
forces also vary with time $t$, these functions will likewise contain
time $t$. I shall assume that these functions are known, since the
acting forces must be included among the known quantities, whether
they depend only on the variables $x$, $y$, $z$ or also on time $t$.\\

{\bf 8}.~~~Let $r$ now express the heat at the point Z or that other
property which, in addition to the density, influences the elasticity
in the case of a compressible fluid. The quantity $r$ must also be
considered to be a function of the three variables $x$, $y$, $z$ and
time $t$, since it might vary with time $t$ at the same point Z in
space. Thus, this function may be regarded as being
known.\footnote{Euler is confining himself to the consideration of
fluid motion in a given temperature field.} Moreover, let the present
density of the fluid particle located at Z be equal to $q$. As the
unit of density I shall take the density of a certain homogeneous
substance which I shall use to measure pressures in terms of heights,
as explained at greater length in my memoir on the equilibrium of
fluids.\footnote{Clearly, for Euler the density $q$ is
non-dimensional, being divided by the constant density $\rho_0$ of a
certain auxiliary fluid: $q=\rho/\rho_0$. Euler defines the pressure
in the fluid as the height $p$ of a column of this same homogeneous
auxiliary fluid. Thus, for Euler pressure is measured by a quantity
with the dimension of length -- the ratio of the acting pressure to
the constant quantity $\rho_0 g$ (where $g$ is the acceleration of
gravity). For further details see Euler, 1755a.} Let, moreover, the
present value of the fluid pressure at the point Z, expressed in terms
of height, be equal to $p$ , which will thus also denote the
elasticity.  Since the nature of the fluid is assumed to be known, we
will know the relation between the height $p$ and the quantities $q$
and $r$.\footnote{That is, the ``equation of state'' of the moving
medium is assumed to be known.}  Thus, $p$ and $q$ will likewise be
functions, albeit unknown, of the four variables $x$, $y$, $z$ and
$t$; however if the fluid is not compressible,\footnote{The 1757 printed version of
  the memoir has ``not incompressible'' [\textit{pas incompressible}], but a
   handwritten copy of the manuscript dated 1755, henceforth cited as Euler,
  1755c 
has ``not
  compressible''[\textit{pas compressible}] which is obviously the correct form.
} the pressure $p$ will be
independent of the density $q$ and the other property [\textit{qualit\'e}] $r$
will not enter into consideration at all.\\

{\bf 9}.~~~Finally, whatever the motion corresponding at a given time to
the fluid element located at the point Z, it too can be decomposed in
the directions ZP, ZQ and ZR parallel to our three axes. Thus, let $u$,
$v$ and $w$ be the velocities of this motion decomposed in the three
directions ZP, ZQ and ZR. It is then obvious that these three
quantities must also be considered to be functions of the four
variables $x$, $y$, $z$ and $t$. Indeed, having found  the nature of
these 
functions,  if the time $t$ is assumed to be constant, then by
varying  the coordinates $x$, $y$ and $z$ the three velocities $u$, $v$
and $w$ and hence the true motion imparted to each element of the fluid
at a given time will be known. If, the coordinates $x$, $y$ and $z$
are assumed to be constant and only the time $t$ is  considered to be
variable, we shall find the motion not of some particular element of
the fluid but of all the elements that pass successively through the
same point Z; in other words, at each moment of time the motion of
that fluid element which is then located at the point Z will be
known.\\

{\bf 10}.~~~Let us consider what path will be described by a fluid element
now at Z during the infinitely small\footnote{The differential operator $\1$,
now denoted using roman fonts, was at the time of Euler italicized; we shall
follow his usage.
}  time $dt$; or the point at which it will be an instant
later.\footnote{The intuitive derivation of the equations of motion and
continuity of an ideal (inviscid and non-heat-conducting) compressible fluid
proposed by Euler is valid provided that the functions in question have
bounded derivatives, up to and including the second. The modern derivation of
these equations, based on the integral laws of conservation of mass and
momentum of the fluid particles and the use of the Gauss theorem, is free of
this limitation.} If we express the distance as the product of velocity and time,
a fluid element currently at Z will travel a distance $udt$ in the direction
ZP, a distance $vdt$ in the direction ZQ and a distance $wdt$ in the direction
ZR. Therefore, if we set
$$
{\rm ZP} =u dt, \quad {\rm ZQ}  = vdt, \quad {\rm and}\quad {\rm ZR} =
wdt
$$ and from these three sides complete the construction of the parallelepiped,
then the corner opposite the point Z will represent the point at which the
fluid element in question will be after the time $dt$ and the diagonal of the
parallelepiped, which is equal to $dt\sqrt(uu + vv + ww)$ will give the true
path described.\footnote{In the 1757 printed version, which we here follow, we
usually find the old notation $xx$ rather than $x^2$ for the square of the
quantity $x$ and $\sqrt(\ldots)$ rather than $\sqrt{\ldots}$ for the square
root of an expression. The manuscript Euler, 1755c, which is not in Euler's
hand, uses modern notation.
} Consequently, the velocity of this true motion
will be equal to $\sqrt(uu + vv + ww)$ and the direction can easily be
determined from the sides of the parallelepiped since it will be inclined to
the plane AOB at an angle whose sine is equal to
$$
\frac{w}{\sqrt(uu+vv+ww)}\;,
$$ 
to the plane AOC at an angle whose sine is equal to 
$$
\frac{v}{\sqrt(uu+vv+ww)}\;,
$$ 
and, finally, to
the plane BOC at an angle whose sine is equal to 
$$
\frac{u}{\sqrt(uu+vv+ww)}\;.
$$\\ 

{\bf 11}.~~~Having determined the motion of a fluid element which at a given
instant is located at the point Z, let us now also examine that of
some other infinitely close element located at the point $z$ with the
coordinates $x + dx$, $y  + dy$ and $z + dz$. The three velocities of
this element in the direction of the three axes can thus be expressed
by $u$, $v$, $w$ after substituting in those quantities $x + dx$, $y +
dy$ and $z + dz$ or after adding to them their differentials while
assuming the time $t$ to be constant. Thus, when $x + dx$ is substituted
for $x$, the increments of $u$, $v$ and $w$ will be:\footnote{Rather
than the now customary notation for partial derivatives using the
symbol $\partial$, Euler employs only the symbol $d$ but encloses the
expressions for partial derivatives in round brackets.}
$$
dx\left( \frac{du}{dx}\right), \quad dx\left(
\frac{dv}{dx}\right),\quad dx\left( \frac{dw}{dx}\right)\;,
$$
and when $y + dy$ is substituted for $y$, the
increments will be: 
$$
dy\left( \frac{du}{dy}\right), \quad dy\left(
\frac{dv}{dy}\right),\quad dy\left( \frac{dw}{dy}\right)\;,
$$
and the same will apply to the variation of
$z$. Then, the three velocities of the fluid element currently located
at $z$ will be:
\par  in the direction OA 
$$
u+dx \left(\frac{du}{dx}\right) +dy\left(\frac{du}{dy}\right) +dz\left(\frac{du}{dz}\right)\;,
$$
\par  in the direction OB 
$$
v+dx \left(\frac{dv}{dx}\right) +dy\left(\frac{dv}{dy}\right) +dz\left(\frac{dv}{dz}\right)\;,
$$
\par  in the direction OC
$$
w+dx \left(\frac{dw}{dx}\right) +dy\left(\frac{dw}{dy}\right) +dz\left(\frac{dw}{dz}\right)\;.
$$\\

{\bf 12}.~~~These are the velocities corresponding to a fluid element at
the point $z$, which is infinitely close to the point Z and whose
position is determined by the three coordinates $x +dx$, $y + dy$ and $z +
dz$. Thus, if we choose a point Z (Fig.~2) such that only $x$ changes by
$dx$, the other two coordinates $y$ and $z$ remaining the same as for the
point Z, the three velocities of the fluid element located at this
point $z$ will be: 
$$
u+dx\left(\frac{du}{dx}\right),\quad v+dx\left(\frac{dv}{dx}\right),\quad 
w+dx\left(\frac{dw}{dx}\right)\;.
$$
These velocities will transport the element in the
time $dt$ to another point $z'$ whose position must be determined
relative to the point $\mathrm{Z'}$, namely the point to which the fluid element
which was at Z is transported in the same time $dt$ and whose position
was determined above (see \S~10). For determining this point $z'$, I
note that if the velocities of the point $z$ were exactly the same as
those of Z, then the point $z'$ would fall at the point
$p$,\footnote{Euler frequently uses the same notation for different
  quantities. Thus, both here 
and later on, the letters $p$ and $q$, which in this article are 
mainly employed to denote pressure and density, are used to denote
certain 
auxiliary points.} such that the
distance ${\rm Z}'p$ would be equal and parallel to the distance Z$z$. Since, by
hypothesis, Z$z$ is parallel to the OA axis and equal to $dx$, the segment
${\rm Z}'p$ will also be equal to $dx$ and parallel to the OA axis.
\begin{figure}[!h]
  \includegraphics[scale=0.8,angle=-1]{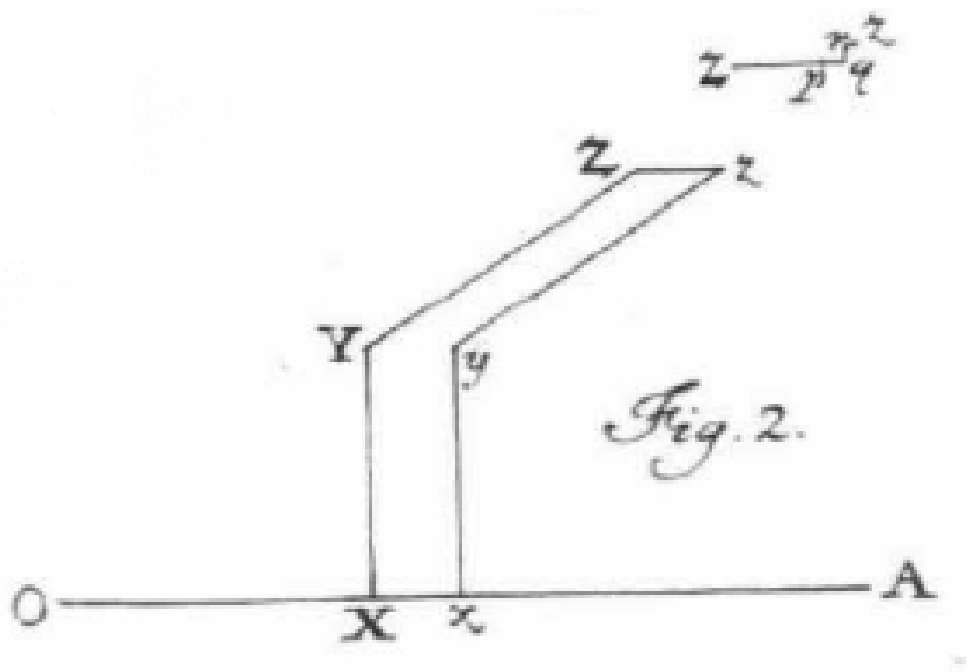}%
\end{figure}
\\

{\bf 13}.~~~Now, since the velocity along OA is not $u$ but $u +
dx\left(\frac{du}{dx}\right)$, this velocity increment will transport
the element in question from $p$ to $q$ in the direction ${\rm Z}'p$, such
that $pq =dtdx\left(\frac{du}{dx}\right)$: this element would thus be
at $q$, if the other two velocities were equal to $v$ and $w$. However,
since the velocity along the OB axis is $v +
dx\left(\frac{dv}{dx}\right)$, this increment will transport our
element from $q$ to $r$, through the distance $qr =
dtdx\left(\frac{dv}{dx}\right)$, and parallel to the axis OB. Finally,
the increment $dx\left(\frac{dw}{dx}\right)$ of the velocity $w$ will
transport the element from $r$ to $z'$ through the infinitesimal
distance [\textit{particule d'espace}]\footnote{The 1757 printed version of the
memoir has ``through the particle'' [\textit{par la particule}], but Euler
1755c has ``through the
particle of distance'' [\textit{par la particule d'espace}].
}  $rz' =
dtdx\left(\frac{dw}{dx}\right)$, and parallel to the third axis
OC. From this I conclude that the fluid element which occupied the
small linear segment Z$z$ would be transported in the time $dt$ to the
segment ${\rm Z}'z'$, inclined at an infinitely small angle to the OA axis,
whose length by virtue of the fact that ${\rm Z'}q=
dx\left(1+dt\left(\frac{du}{dx}\right)\right)$ will be
$$
dx \,{\raise 1.ex\hbox{$\sqrt{}$}} 
\left(\left( 1+dt\left(\frac{du}{dx}\right)\right)^2
+dt^2\left(\frac{dv}{dx}\right)^2 +dt^2\left(\frac{dw}{dx}\right)^2
\right)\;.
$$
Thus, neglecting the terms that contain the square of $dt$, the
length 
${\rm Z}'z'$ 
will not differ from ${\rm Z}'q$ and we  shall have: ${\rm Z'}z' =
dx\left(1+dt\left(\frac{du}{dx}\right)\right)$. For the inclination of this line to the OA axis, it will suffice to
 note that it is an infinitely small quantity of 
the first order and can be expressed as $\alpha dt$.
\begin{figure}[!h]
  \includegraphics[scale=0.5,angle=-1]{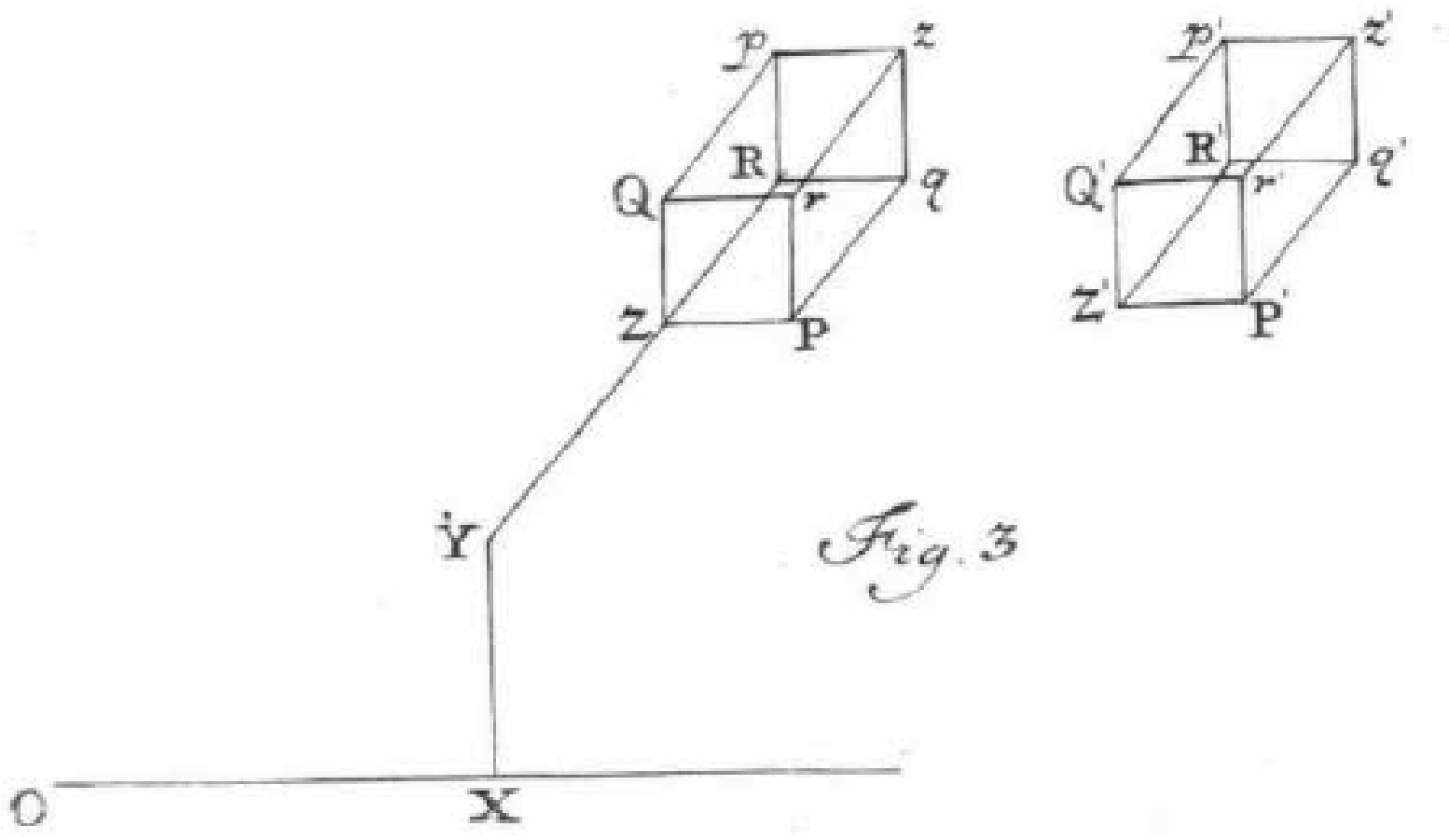}%
\end{figure}
\\ 

{\bf 14}.~~~If the small segment Z$z$ had been taken equal to $dy$ and
parallel to the OB axis, by the same reasoning it could have been
shown that the fluid which occupied that segment would have been
transported to another segment ${\rm Z'}z'
=dy\left(1+dt\left(\frac{dv}{dy}\right)\right)$, and which would have been inclined to the
OB axis at an infinitely small angle. And if we had taken the segment 
${\rm Z}z =dz$, and parallel to the third axis OC, the fluid which
occupied it would have have been transported to another segment ${\rm
  Z'}z' =dz\left(1+dt\left(\frac{dw}{dz}\right)\right)$, and which
would have been inclined to the OC axis at an  infitely small angle.
Thus, if we consider a rectangular
parallelepiped ZPQR$zpqr$ (Fig.~3) formed by the three sides ${\rm ZP}=dx$,
${\rm ZQ}=dy$, ${\rm ZR}=dz$, the fluid occupying that volume would be transported in
the time $dt$ to fill a volume ${\rm Z}'{\rm P}'{\rm Q}'{\rm R}'z'p'q'r'$ differing infinitely
slightly from a rectangular parallelepiped whose three sides would be
\begin{eqnarray}
&&{\rm Z'P'} =dx\left(1+dt\left(\frac{du}{dx}\right)\right)\nonumber;\\
&&{\rm Z'Q'} =dy\left(1+dt\left(\frac{dv}{dy}\right)\right)\nonumber;\\
&&{\rm Z'R'}=dz\left(1+dt\left(\frac{dw}{dz}\right)\right)\nonumber\;.
\end{eqnarray}
Since the sides ZP, ZQ, ZR go over into ${\rm Z}'{\rm P}'$, ${\rm
  Z}'{\rm Q}'$, ${\rm Z}'{\rm R}'$, there is no
doubt that the fluid contained in the first volume will be transported into
the other in the time $dt$.\\  

{\bf 15}.~~~We can now judge whether'the volume of fluid occupying the
parallelepiped Z$z$ has increased or decreased in the time $dt$. For this
we need only to find the volume or the capacity of each of these two
solids. Since the first is a parallelepiped formed by the sides $dx$, $dy$,
$dz$, its volume is equal to $dxdydz$. As for the other, whose plane
angles differ infinitely slightly from a right angle, I note that its
volume can also be found by multiplying its three sides, since the
error due to the infinitesimal distortion of the angles will
enter into terms which contain the square of the time element $dt$ and
can therefore be neglected. Thus, the volume ${\rm Z}'z'$ can be represented
by the expression: 
$$
dxdydz\left(1+dt\left(\frac{du}{dx}\right)+dt\left(\frac{dv}{dy}\right) +dt\left(\frac{dw}{dz}\right)\right)\;.
$$
Anyone still harboring doubts about the
reasonableness of this conclusion need only consult my Latin paper
\textit{Principia motus fluidorurn} in which I calculate this volume without
neglecting anything.\footnote{See Euler, 1756--1757. This memoir was
  originally entitled \textit{De motu fluidorum in genere}, but the
  final title was already used here.
}\\

{\bf 16}.~~~Thus, if the fluid is not compressible, these two volumes
should be equal, since the mass occupying the volume Z$z$ would not fit
into either a larger or a smaller volume. However, since I propose to
examine the problem in the most general possible form and have denoted
the density at Z by $q$, considering $q$ to be a function of the three
coordinates and time, I note that to find the density at ${\rm Z}'$ it will
first be necessary to increase the time $t$ by its differential $dt$;
then, as the point ${\rm Z}'$ is different from Z, the quantities $x$, $y$, $z$ will
have to be increased by the small increments $udt$, $vdt$, $wdt$; whence
the density at ${\rm Z}'$ will be:
$$
q+dt\left(\frac{dq}{dt}\right)+udt\left(\frac{dq}{dx}\right)+vdt\left(\frac{dq}{dy}\right)+wdt\left(\frac{dq}{dz}\right)
$$
 and since the density is inversely
proportional to the volume, this quantity will be to $q$ as $dxdydz$
 to
$$
dxdydz\left(1+dt\left(\frac{du}{dx}\right)+dt\left(\frac{dv}{dy}\right)
+dt\left(\frac{dw}{dz}\right)\right)\,.
$$
Thus, dividing by $dt$, we find that consideration of the density leads
to the following equation:
\begin{eqnarray}
\left(\frac{dq}{dt}\right) &+&u\left(\frac{dq}{dx}\right)
  +v\left(\frac{dq}{dy}\right) +w\left(\frac{dq}{dz}\right)\nonumber
  \\
&+& q\left(\frac{du}{dx}\right)+ q\left(\frac{dv}{dy}\right)
+ q\left(\frac{dw}{dz}\right)=0\;.\nonumber
\end{eqnarray}
\\ 

{\bf 17}.~~~Here, then, is a very remarkable condition which already
establishes a certain relation between the three velocities $u$, $v$ and
$w$ and the fluid density $q$. Now this equation can be reduced to a
simpler form.\footnote{In Euler's subsequent exposition the use of round brackets goes beyond the scope of simple partial derivative notation, but the meaning 
of the operations is still clear, in Euler's notation $d.qu=d(qu)$,
etc.} 
Thus, $u\left(\frac{dq}{dx}\right)$ is no different from
$\left(u\frac{dq}{dx}\right)$  since this
form of expression must be taken to mean that in differentiating $q$
only the quantity $x$ is taken to be a variable, and similarly
$q\left(\frac{du}{dx}\right)=\left(q\frac{du}{dx}\right)$;
from which it follows that
$$
q\left(\frac{du}{dx}\right) + u\left(\frac{dq}{dx}\right)
=\left(\frac{udq+qdu}{dx}\right) = \left(\frac{d.qu}{dx}\right)\;,
$$ the differential of the product $qu$ being so understood that only
the quantity $x$ is regarded as a variable. Accordingly, the equation
obtained can be reduced to the following:
$$
\left(\frac{dq}{dt}\right) +\left(\frac{d.qu}{dx}\right)+\left(\frac{d.qv}{dy}\right)+\left(\frac{d.qw}{dz}\right)=0\;.
$$
If the fluid was not
compressible, the density $q$ would be the same at both Z and ${\rm Z}'$ and for
this case we would have the equation:
$$
\left(\frac{du}{dx}\right)+\left(\frac{dv}{dy}\right)+\left(\frac{dw}{dz}\right)=0\;,
$$
 which is also that on which I
based my Latin memoir mentioned above.\footnote{See Euler, 1756--1757.}\\

{\bf 18}.~~~This equation, obtained by considering the continuity of
the fluid, already contains a certain relation which must exist
between the quantities $u$, $v$, $w$ and $q$. The other relations must
be obtained by considering the forces to which each fluid particle is
subjected. Thus, in addition to the accclerative
forces\footnote{Concerning the concept of ``accelerative'' (body)
forces, see footnote~\ref{fn:newton}.}  P, Q, R, which act on the
fluid at Z, the fluid is also subjected to the pressure
[\textit{pression}] exerted from all sides on the fluid element
contained at Z. Combining these two forces, we obtain three
accelerative forces in the direction of the three axes.  Since the
accelerations themselves can be determined by considering the
velocities $u$, $v$ and $w$, we can derive three equations which,
together with that which we have just found, will contain everything
that relates to the motion of fluids, so that we shall then have the
general and complete laws of the entire science of fluid motion.\\

{\bf 19}.~~~In order to find the accelerations undergone by a fluid element
at Z, we need only compare the velocities $u$, $v$, $w$ which currently
correspond to the point Z with the velocities corresponding to the
point ${\rm Z}'$ after the lapse of the time $dt$. Thus, a double change takes
place: with respect to the coordinates $x$, $y$, $z$, which receive the
increments $u dt$, $vdt$, $wdt$, as well as with respect to time, which increases
by $dt$. Hence it follows that the three velocities at the point ${\rm
  Z}'$ are:
\par in the direction OA 
$$
\makebox[1.6ex][l]{$u$}+dt\left(\frac{d\makebox[1.6ex][c]{$u$}}{dt}\right)+udt\left(\frac{du}{dx}\right)+vdt 
\left(\frac{d\makebox[1.6ex][c]{$u$}}{dy}\right)+wdt\left(\frac{d\makebox[1.6ex][c]{$u$}}{dz}\right)\;,
$$
\par in the direction OB
$$
\makebox[1.6ex][l]{$v$}+dt\left(\frac{dv}{dt}\right)+udt\left(\frac{d\makebox[1.6ex][c]{$v$}}{dx}\right)+vdt 
\left(\frac{d\makebox[1.6ex][c]{$v$}}{dy}\right)+wdt\left(\frac{d\makebox[1.6ex][c]{$v$}}{dz}\right)\;,
$$
\par in the direction OC
$$
w+dt\left(\frac{dw}{dt}\right)+udt\left(\frac{dw}{dx}\right)+vdt 
\left(\frac{dw}{dy}\right)+wdt\left(\frac{dw}{dz}\right)\;,
$$
and hence the accelerations, expressed in terms of the velocity 
increments divided by the time element $dt$, will be:
\par in the direction OA 
$$
\left(\frac{du}{dt}\right)+u\left(\frac{du}{dx}\right)+v 
\left(\frac{du}{dy}\right)+w\left(\frac{du}{dz}\right)\;,
$$
\par in the direction OB
$$
\left(\frac{dv}{dt}\right)+u\left(\frac{dv}{dx}\right)+v 
\left(\frac{dv}{dy}\right)+w\left(\frac{dv}{dz}\right)\;,
$$
\par in the direction OC
$$
\left(\frac{dw}{dt}\right)+u\left(\frac{dw}{dx}\right)+v 
\left(\frac{dw}{dy}\right)+w\left(\frac{dw}{dz}\right)\;.
$$\\

{\bf 20}.~~~We will now seek the accelerative forces acting in these same
 directions due to the pressure exerted by the fluid on the
 parallelepiped Z$z$, whose volume is equal to $dxdydz$, the mass of the
 fluid occupying that volume thus being equal to $qdxdydz$. Since the
 pressure at the point Z is expressed in terms of the height $p$, the
 motive force acting on the face ZQR$p$ is equal to $pdxdydz$. For the
 opposite face $zqr$P with the area $dydz$, the height $p$ is increased by
 its differential $dx\left(\frac{dp}{dx}\right)$, obtained on the
 assumption 
that only $x$ is
 variable. Accordingly, this fluid mass Z$z$ is driven in the direction
 AO by the motive force $dxdydz\left(\frac{dp}{dx}\right)$ or by the
 accelerative force $\frac{1}{q}\left(\frac{dp}{dx}\right)$.
 Similarly, we find that the fluid mass Z$z$ is subjected to
 the action of the accelerative force 
$\frac{1}{q}\left(\frac{dp}{dy}\right)$ in the direction BO
 and to that of the accelerative force 
$\frac{1}{q}\left(\frac{dp}{dz}\right)$ in the direction
 CO. To these forces we add the given forces P, Q, R, and the  total
 accelerative forces will be:
\begin{eqnarray}
 &&\hbox{in the direction OA:}\quad {\rm P}-
  \frac{1}{q}\left(\frac{dp}{dx}\right)\nonumber\\ 
&&\hbox{in the direction OB:}\quad {\rm Q}-
  \frac{1}{q}\left(\frac{dp}{dy}\right)\nonumber\\ 
&&\hbox{in the direction OC:}\quad {\rm R}-
  \frac{1}{q}\left(\frac{dp}{dz}\right)\nonumber\;.
\end{eqnarray}\\

{\bf 21}.~~~Thus, it only remains to equate these accelerative forces with
the actual accelerations which we have just found.  We then obtain
the following three equations:\footnote{Despite the outward resemblance between Euler's
equations and modern notation, they have been written here in
dimensionless form. As mentioned above, the pressure $p$ is measured as
the ratio of the acting pressure to the specific weight $\rho_0g$ of
a certain homogeneous auxiliary fluid, the density $q$ is
dimensionless ($q= \rho/\rho_0$),  the components of the body forces have been
divided by the acceleration of gravity $g$, the transition from the
Eulerian velocities $u$, $v$, $w$ to the real velocities $U$, $V$, $W$
is effected
by means of a transformation of the form $u \mapsto  U/\sqrt{g}$ and the transition
from Eulerian time to real time by means of the transformation $t
\mapsto T\sqrt{g}$. (For further details concerning Euler's system of physical
units, see Mikhailov, 1999.)}
\begin{eqnarray}
{\rm P}-
  \frac{1}{q}\left(\frac{dp}{dx}\right)\!\!&=&\!\!\left(\frac{d\,u}{dt}\right)+u\left(\frac{d\,u}{dx}\right)+v 
\left(\frac{d\,u}{dy}\right)+w\left(\frac{d\,u}{dz}\right)\nonumber\\
{\rm Q}-
  \frac{1}{q}\left(\frac{dp}{dy}\right)\!\!&=&\!\!\left(\frac{d\,v}{dt}\right)+u\left(\frac{d\,v}{dx}\right)+v 
\left(\frac{d\,v}{dy}\right)+w\left(\frac{d\,v}{dz}\right)\nonumber\\
{\rm R}-
  \frac{1}{q}\left(\frac{dp}{dz}\right)\!\!&=&\!\!\left(\frac{dw}{dt}\right)+u\left(\frac{dw}{dx}\right)+v 
\left(\frac{dw}{dy}\right)+w\left(\frac{dw}{dz}\right)\nonumber.
\end{eqnarray}
If we add to these three
equations, first, that obtained from considering the continuity of the
fluid, namely
$$
\left(\frac{dq}{dt}\right) +\left(\frac{d.qu}{dx}\right)+\left(\frac{d.qv}{dy}\right)+\left(\frac{d.qw}{dz}\right)=0\;,
$$ 
and then the equation\footnote{What we now call the equation of state.}  which gives the relation between
the elasticity  $p$, the density $q$ and the other property $r$
which, in addition to the density $q$ influences the elasticity $p$, we shall
have five equations encompassing the entire Theory of the motion of
fluids.\\

{\bf 22}.~~~Whatever be the nature of the forces P, Q, R, provided that they
are real, it should be noted that ${\rm P}dx + {\rm Q}dy + {\rm R}dz$ 
is always a total [\textit{r\'eel}] differential of a certain finite and
determinate 
quantity,\footnote{Euler is thinking here of real body forces
  possessing a potential (more correctly, a force function). By
  ``finite'' quantities (functions) 
Euler 
means quantities that do not contain differentials.}
assuming the three coordinates $x$, $y$ and $z$ to be variables. Thus, we
will always have:
$$
\left(\frac{d{\rm P}}{dy}\right)=\left(\frac{d{\rm Q}}{dx}\right); \!\!\quad\left(\frac{d{\rm P}}{dz}\right)=\left(\frac{d{\rm R}}{dx}\right);\!\!\quad
\left(\frac{d{\rm Q}}{dz}\right)=\left(\frac{d{\rm R}}{dy}\right)\;,
$$
 and if we set this finite quantity equal to S, then,
 we have $$
d{\rm S}={\rm P}dr + {\rm Q}dy + {\rm R}dz\;,
$$
assuming the time $t$ to be constant for
the case in which the forces P, Q, R also vary with time at the same
points.  The quantity S expresses what I shall call the effort
[\textit{l'effort}] of the acting forces\footnote{\label{fn:effort}Euler's
  ``effort'' is equivalent to the modern notion of potential.}
and is equal to the sum of the
integrals of each force multiplied by the elementary interval in the
direction of that force or by the small distance through which it
would drag a body subjected to its action. This notion of effort is of
the utmost importance for the entire theory of both equilibrium and
motion, since it makes it possible to see that the sum of all the
efforts is always a maximum or a minimum. This excellent property fits
in admirably with the splendid principle of least action whose
discovery we owe to our illustrious President, Mr. de
Maupertuis.\footnote{Maupertuis was president of the Berlin Academy at
  the time.}\\

{\bf 23}.~~~The equations just obtained contain four variables $x$, $y$,
$z$ and $t$ which are absolutely independent of each other since the
variability of the first three extends to all elements of the fluid
and that of the fourth to all times. Therefore, for the equations to
continue to hold, the other variables $u$, $v$, $w$, $p$ and $q$ must
be certain functions of the former. For although a differential
equation with two variables\footnote{Here, by variable Euler understands both independent variables and their functions.} is always possible,\footnote{We would now
say ``soluble''.
} we know that a differential equation containing
three or more variables is possible only under certain conditions, by
virtue of which a certain relationship must exist between the terms of
the equation. Therefore, before we can begin solving the equations, we
need to know what sort of functions of $x$, $y$, $z$ and $t$ must be
used to express the values of $u$, $v$, $w$, $p$ and $q$ in order for
these same equations to be possible.\\

{\bf 24}.~~~We now multiply the first of the three equations obtained by
$dx$, the second by $dy$ and the third by $dz$, and since 
$dx\left(\frac{dp}{dx}\right)+dy\left(\frac{dp}{dy}\right)+dz\left(\frac{dp}{dz}\right)$
represents the
differential of $p$, assuming only time $t$ to be constant, we
obtain\footnote{The first term on the r.h.s. is correct in the manuscript 
Euler, 1755c but misprinted as  $dz\left(\frac{du}{dt}\right)$ in the printed version.}
\begin{eqnarray}
d{\rm S}\!&-&\!\frac{dp}{q} = \nonumber\\
&+&\!\!dx\left(\frac{d\makebox[1.5ex][r]{$u$}}{dt}\right)+udx\left(\frac{d\makebox[1.5ex][r]{$u$}}{dx}\right)+vdx\left(\frac{d\makebox[1.5ex][r]{$u$}}{dy}\right)+wdx\left(\frac{d\,u}{dz}\right)\nonumber\\
&+&\!\!dy\left(\frac{d\makebox[1.7ex][r]{$v$}}{dt}\right)+udy\left(\frac{d\makebox[1.7ex][r]{$v$}}{dx}\right)+vdy\left(\frac{d\makebox[1.7ex][r]{$v$}}{dy}\right)+wdy\left(\frac{d\makebox[1.7ex][r]{$v$}}{dz}\right)\nonumber\\
&+&\!\!dz\left(\frac{dw}{dt}\right)+udz\left(\frac{dw}{dx}\right)+vdz\left(\frac{dw}{dy}\right)+wdz\left(\frac{dw}{dz}\right)\nonumber.
\end{eqnarray}
 It is
now a question of finding the integral of this equation in which time
is assumed to be constant. It should be noted that this single
equation contains the three equations of which it is composed and
that as soon as it is satisfied the conditions of all three will be
fulfilled. Thus, if the expression $d{\rm S} - \frac{dp}{q}$ is equal
to the 
three
lines, where $x$, $y$ and $z$ are variables, the portion of $d{\rm S} -
\frac{dp}{q}$ due to
the variability of $x$ alone, namely ${\rm P}dx
-\frac{dx}{q}\left(\frac{dp}{dx}\right)$ must necessarily be equal to the
first line, and similarly for the other two. The terms 
$\left(\frac{du}{dt}\right)$, $\left(\frac{dv}{dt}\right)$, and 
$\left(\frac{dw}{dt}\right)$, found by
assuming the variability of time $t$, since they denote certain finite
functions, do not prevent time $t$ from now being taken to be
constant.\\

{\bf 25}.~~~Suppose that this equation has already been solved and the
quantities $u$, $v$, $w$, $q$ and $p$ have been found as certain finite
functions of $x$, $y$, $z$ and $t$. The substitution of these functions in
the differential equation, with time $t$ assumed constant, yields an
identity. Since after this substitution we will have three types of
terms, the first associated with $dx$, the second with $dy$ and the third
with $dz$, the identity leads us to three equations whence it is dear
that although only one differential equation is being considered, it
actually has the force of three and determines three of our
unknowns. What is also clear is that a differential equation with three
variables, such as ${\rm L}dx + {\rm M}dy + {\rm N}dz =0$, 
cannot be solved unless a
certain relationship exists between the quantities L, M and N.
However, since very little work has yet been done on solving these
three-variable equations, we cannot hope to obtain a complete solution
of our equation until the limits of Analysis have been extended much
further.\\

{\bf 26}.~~~The best approach would therefore be to ponder well on the particular
solutions of our differential equation that we are in a position to obtain, as
this would enable us to judge which path to follow in order to arrive at a
complete solution. I have already pointed out\footnote{Euler,
1756--1757:~\S\S~60--67.} that where the density $q$ is assumed to be constant a
very elegant solution can be obtained when the velocities $u$, $v$ and $w$ arc
such that the differential expression [\textit{formule}] $udx + vdy + wdz$ can be integrated.
Suppose, then, that W is that integral, being any
function of $x$, $y$, $z$ and time $t$, and that its differentiation, also
including  $t$ as a variable, gives 
$$
d{\rm W}=u dx + vdy + wdz + \Pi dt\;.
$$ 
Then the quantities $u$, $v$, $w$ and $\Pi$ will be related as
follows:\footnote{ In modem terminology, the function introduced by Euler $W=
W(x,y, z, t)$ is the velocity potential; here, the equality of the cross
derivatives of $W$ with respect to the coordinates (condition of integrability
of $dW$) is the condition of absence of vorticity.} 
\begin{eqnarray}
\!\!\left(\frac{du}{dy}\right)&=&\left(\frac{d\makebox[1.7ex][r]{$v$}}{dx}\right); \quad\left(\frac{du}{dz}\right)=\left(\frac{dw}{dx}\right);\quad
\left(\frac{d\makebox[1.7ex][r]{$u$}}{dt}\right)=\left(\frac{d\Pi}{dx}\right);\nonumber\\
\!\!\left(\frac{dv}{dz}\right)&=&\left(\frac{dw}{dy}\right); \quad\left(\frac{dv}{dt}\right)=\left(\frac{d\Pi}{dy}\right);\quad
\left(\frac{dw}{dt}\right)=\left(\frac{d\Pi}{dz}\right).\nonumber
\end{eqnarray}
\\

{\bf 27}.~~~Using these equalities, we can reduce our differential
equation to the following form: 
\begin{eqnarray}
d{\rm S}\!&-&\!\frac{dp}{q} = \nonumber\\
&+&\!\!dx\left(\frac{d\Pi}{dx}\right)+udx\left(\frac{d\makebox[1.6ex][c]{$u$}}{dx}\right)+vdx\left(\frac{d\makebox[1.6ex][c]{$u$}}{dy}\right)+wdx\left(\frac{d\makebox[1.6ex][c]{$u$}}{dz}\right)\nonumber\\
&+&\!\!dy\left(\frac{d\Pi}{dy}\right)+udy\left(\frac{d\makebox[1.6ex][c]{$v$}}{dx}\right)+vdy\left(\frac{d\makebox[1.6ex][c]{$v$}}{dy}\right)+wdy\left(\frac{d\makebox[1.6ex][c]{$v$}}{dz}\right)\nonumber\\
&+&\!\!dz\left(\frac{d\Pi}{dz}\right)+udz\left(\frac{dw}{dx}\right)+vdz\left(\frac{dw}{dy}\right)+wdz\left(\frac{dw}{dz}\right)\nonumber.
\end{eqnarray}
Since here time $t$ is assumed to be
constant, using the same hypothesis we will have 
\begin{eqnarray}
&&dx\left(\frac{d\Pi}{dx}\right)+dy\left(\frac{d\Pi}{dy}\right)+
dz\left(\frac{d\Pi}{dz}\right)= d\Pi \nonumber \\
&&dx\left(\frac{d\makebox[1.7ex][c]{$u$}}{dx}\right)+dy\left(\frac{d\makebox[1.7ex][c]{$u$}}{dy}\right)+
dz\left(\frac{d\makebox[1.7ex][c]{$u$}}{dz}\right) = d\makebox[1.7ex][c]{$u$}\nonumber \\[1.9ex]
&& .............................................................\nonumber
\end{eqnarray}
Thus, our equation will become 
$$
d{\rm S}-\frac{dp}{q} =d\Pi =udu+vdv+wdw\;,
$$
or
$$
dp =q\left(d{\rm S}-d\Pi -udu -vdv-wdw\right)\;.
$$
Hence, if the density of the fluid is everywhere the same, or $q=g$, 
as a result of integration we obtain:\footnote{The subsequent equation, which generalizes the Bernoulli integral, is usually
 associated with the names of Cauchy and Lagrange.}
$$
p =g\left({\rm C}+{\rm S} -\Pi -{\scriptstyle \frac{1}{2}}uu -{\scriptstyle \frac{1}{2}}vv-{\scriptstyle \frac{1}{2}}ww\right)\;.
$$\\

{\bf 28}.~~~For brevity, let us set 
$$
{\rm C}+ {\rm S} - \Pi  -{\scriptstyle \frac{1}{2}}uu -{\scriptstyle
  \frac{1}{2}}vv-{\scriptstyle \frac{1}{2}}ww = {\rm V}\;,
$$ 
where it should be noted that the constant C may well contain the
time $t$, since it is considered to be constant in this integration
and, as $dp=qd{\rm V}$, it is clear that the hypothesis 
$$
d{\rm W}=u dx + vdy + wdz + \Pi dt\;,
$$
also makes our differential equation possible, when the elasticity
$p$ depends in any way on the density $q$ only or $q$ is any function
of $p$. It will also become possible if the fluid is not compressible but the
density $q$ varies in such a way that it is an arbitrary function of
the quantity V. And in general, if the elasticity $p$ depends both on the
density $q$ and on some other quantity represented by the letter $r$, the
hypothesis may also be satisfied provided that $r$ is a function of
V. In all these cases, for the motion to exist under this hypothesis
it is also necessary for the following condition to be satisfied:
$$
\left(\frac{dq}{dt}\right) +\left(\frac{d.qu}{dx}\right)+\left(\frac{d.qv}{dy}\right)+\left(\frac{d.qw}{dz}\right)=0\;.
$$\\

{\bf 29}.~~~This hypothesis is so general that it seems that there is not
a single case that is not included and hence that, generally speaking,
the equation $dp= qd{\rm V}$, together with the other equations which
present hardly any difficulty, incorporates all the foundations of
the Theory of the motion of fluids. Thus, I concerned myself
exclusively with this case in my Latin memoir on the laws of fluid
motion\footnote{See Euler, 1756--1757.} in which I considered
incompressible fluids only and showed that all the cases previously
considered, in which the fluid moves through pipes of arbitrary shapes,
are contained in this supposition and that the velocities
$u$, $v$ and $w$ are always such that the differential expression $u dx +
vdy + wdz$  is integrable. However, I have since noted that there are
also cases, even when the fluid is incompressible and everywhere
homogeneous, in which this condition does not hold, which is enough to
convince me that the solution I have just given is only a particular
one.\footnote{Here, Euler recognizes that his previous memoir on
fluid motion was too restricted, in so far as it ignored what
we now call vorticity.
}\\

{\bf 30}.~~~To give an example of a real motion which would be perfectly
 consistent with all the equations that follow from the laws of
 Mechanics, but without the expression $udx + vdy + wdz$ being
 integrable, 
let us assume that the fluid is incompressible and everywhere
 homogeneous, i.e.\ that $q$ is constant and equal to $g$, and that there
 are no forces acting on the fluid, so that ${\rm P}=0$, ${\rm Q}=0$ 
and ${\rm R}=0$. Then,
 let $w=0$, $v={\rm Z}x$ and $u = -{\rm Z}y$, where Z denotes any
 function of $\sqrt(xx+yy)$.
 It is now obvious that the expression $u dx + vdy + wdz$,
 which takes the form $-{\rm Z}ydx + {\rm Z}xdy$, is integrable only
 in 
the case ${\rm Z} = \frac{1}{xx+yy}$.
 However, these values\footnote{The corresponding values of $u$, $v$ 
and $w$.} satisfy all our
 formulas so that the possibility of this motion cannot be
 questioned.  Since Z is a function of  $\sqrt(xx+yy)$, its differential
 will have the form $d{\rm Z}={\rm L}x dx + {\rm L} ydy$, where L will
 again be any function of $\sqrt(xx+yy)$.\\

{\bf 31}.~~~Using these values of $u$, $v$ and $w$, we obtain:
$$
\begin{array}{llr}
\ds\left(\frac{du}{dt}\right)=0;& \ds\left(\frac{dv}{dt}\right)=0;&\ds\left(\frac{dw}{dt}\right)=0;\\[1.9ex]
\ds\left(\frac{du}{dx}\right)=-{\rm L}xy;&\ds
\left(\frac{dv}{dx}\right)={\rm Z}+{\rm L}xx;&\ds\left(\frac{dw}{dx}\right)=0;\\[1.9ex]
\ds\left(\frac{du}{dy}\right)=-{\rm Z}-{\rm L}yy;& \ds\left(\frac{dv}{dy}\right)={\rm L}xy;&\ds \left(\frac{dw}{dy}\right)=0;\\[1.9ex]
\ds\left(\frac{du}{dz}\right) =0;&\ds
\left(\frac{dv}{dz}\right)=0;&\ds
\left(\frac{dw}{dz}\right)=0;
\end{array}
$$ 
and since $d{\rm S}=O$, assuming time $t$ to be constant, we have the
following differential equation: 
\begin{eqnarray}
\frac{dp}{g}&=&{\rm L}{\rm Z}xyydx-{\rm Z}{\rm Z}xdx\nonumber\\
&-&{\rm L}{\rm Z}xyydx-{\rm Z}{\rm
  Z}ydy- {\rm L}{\rm Z}xxydy+{\rm L}{\rm Z}xxydy\nonumber\\
&=&-{\rm Z}{\rm Z}(xdx+ydy)\nonumber\;.
\end{eqnarray}
Consequently $dp=g{\rm Z}{\rm Z}(xdx+ydy)$,
since Z is assumed to be a function of $\sqrt(xx + yy)$, this equation
will definitely be possible and will yield the integral  $p=g\int{\rm
  Z}{\rm Z}(xdx+ydy)$. We see that the differential equation would
also 
be possible if
the fluid were subjected to the action of certain arbitrary forces P,
Q, R, provided that the expression ${\rm P}dx+{\rm Q}dy+{\rm R}dz$ was a total
[\textit{possible}] differential equal to $d{\rm S}$, since then $p= g{\rm S} +
g\int{\rm   Z}{\rm Z}(xdx+ydy)$.\\

{\bf 32}.~~~As these values $u = -{\rm Z}y$, $v={\rm Z}x$ and $w=0$ 
satisfy our
differential equation, they can also be seen to satisfy the condition
contained in the equation:\footnote{Strictly speaking, it cannot be said that the values of $u$, $v$ and $w$ assumed in \S~30
 also satisfy the equation of motion from \S~31; in reality, this 
equation determines the corresponding pressure $ p =p(s)$ ($s= \sqrt(xx
+ yy)$,  the continuity equation being satisfied irrespective of the
equations  of motion.}
$$
\left(\frac{dq}{dt}\right) +\left(\frac{d.qu}{dx}\right)+\left(\frac{d.qv}{dy}\right)+\left(\frac{d.qw}{dz}\right)=0\;.
$$
 By virtue of the fact that $q=g$, this
equation goes over into
$$
-g{\rm L}xy + g{\rm L}xy=0
$$ which, being an identity, satisfies the required conditions. Thus,
it is quite possible for a fluid to have a motion such that the
velocities of each of its elements are $u= -{\rm Z}y$, $v={\rm }Zx$
and $w =0$, although the differential expression $udx + vdy + wdz$ is
not possible;\footnote{That is a total differential.} this confirms
that there are cases in which fluid motion is possible without this
condition, which seemed general, being fulfilled. Thus, the assumption
that the differential expression $udx + vdy + wdz$ is possible yields
only a particular solution of the equations we have found.\\

\setbox0=\hbox{${\scriptstyle \cup}$}
{\bf 33}.~~~Clearly, the motion corresponding to this case reduces to a
rotational motion about the axis OC and since what has been said
about the axis OC can be applied to any other fixed axis, we may 
conclude that it is possible for a fluid acted upon by any forces whose
effort\footnote{See footnote~\ref{fn:effort}.} is equal to S to have a motion about a fixed axis such that
the rotational velocities are proportional to any function of the
distance to that axis. Thus, if the distance to that axis is denoted
by $s$ and the rotation velocity at that distance by
$\ttaurus$,\footnote{In modern terms Z is the angular velocity at the 
radial distance $s$ and $\ttaurus = {\rm Z}s$ is the tangential velocity.} 
then since $xx+ yy=ss$ and ${\rm Z}{\rm Z}ss=\ttaurus \ttaurus$, the pressure 
there will be expressed by the height $p=g{\rm S}
+g\int\frac{\ttaurus \ttaurus ds}{s}$.
Thus, such a motion, which corresponds to that of a vortex
[\textit{tourbillon}], is just as possible as those contained in the
expression $udx + vdy + wdz$ when the latter is integrable. No doubt
there is an infinity of other motions,  which satisfying our equations,
 are also equally possible.\\

{\bf 34}.~~~Let us now return to our general formulas and, since they are
somewhat too complicated, introduce, for greater conciseness, the
notation:
$$
\begin{array}{lcr}
\ds \left(\frac{d\,u}{dt}\right)+u\left(\frac{d\,u}{dx}\right)+v 
\left(\frac{d\,u}{dy}\right)+w\left(\frac{d\,u}{dz}\right)&\ds =&\ds {\rm X}\nonumber\\[1.9ex]
\ds \left(\frac{d\,v}{dt}\right)+u\left(\frac{d\,v}{dx}\right)+v 
\left(\frac{d\,v}{dy}\right)+w\left(\frac{d\,v}{dz}\right)&\ds =&\ds {\rm Y}\nonumber\\[1.9ex]
\ds \left(\frac{dw}{dt}\right)+u\left(\frac{dw}{dx}\right)+v 
\left(\frac{dw}{dy}\right)+w\left(\frac{dw}{dz}\right)&\ds =&\ds {\rm
  Z}\nonumber\;.
\end{array}
$$ Whatever the nature of the three accelerative forces P, Q and R,
  granted  that\footnote{Here, Euler assumes that all real
  body forces 
have a potential ${\rm S}={\rm S}(x, y, z)$.} $d{\rm S} ={\rm P}dx +
{\rm Q}dy + {\rm R}dz$, the differential equation
$$
\frac{dp}{q} = ({\rm P}-{\rm X})\,dx+ ({\rm Q}-{\rm Y})\,dy+({\rm R}-{\rm
  Z})\,dz\;,
$$
in which $t$ is assumed to be constant
must be satisfied. Moreover, the continuity of the fluid requires that:
$$
\left(\frac{dq}{dt}\right) +\left(\frac{d.qu}{dx}\right)+\left(\frac{d.qv}{dy}\right)+\left(\frac{d.qw}{dz}\right)=0\;.
$$
In whatever manner these two equations are satisfied, there will always be a
motion which can actually take place in the fluid.\\

{\bf 35}.~~~If the fluid is everywhere incompressible and homogeneous,
i.e.\ the density $q$ is constant and equal to $g$, then, clearly, the
differential equation cannot be satisfied unless the differential 
$$
({\rm P}-{\rm X})dx+({\rm Q} -{\rm Y})dy+({\rm R}- {\rm Z})dz\;,
$$
is possible or total, i.e. unless it can be obtained as a result of the
actual differentiation of some finite function of the variables $x$, $y$
and $z$, which may also contain the time $t$, although in the
differentiation 
the latter is assumed to be constant. It is also obvious that
this differential expression must be soluble or total when the fluid
is compressible and the density $q$ is expressed in terms of any
function of the elasticity $p$. In both cases, if we denote by V the
finite quantity whose differential has the form: 
$$
d{\rm V}=({\rm P}-{\rm X})dx+({\rm Q}-{\rm Y})dy+ ({\rm R}-{\rm Z})dz\;,
$$
our differential
equation will yield either $\frac{p}{g}= {\rm V}$ or $\int
\frac{dp}{q} = {\rm V}$. In addition, however, for the
motion to be possible the other condition derived from the continuity
must also be fulfilled.\\

{\bf 36}.~~~If the fluid is not compressible, but its density $q$ is
variable and can be expressed in terms of any function of position,
i.e.\ of the three coordinates $x$, $y$, $z$ and time $t$, it is not sufficient
for the expression
$$
({\rm P}-{\rm X})dx+({\rm Q}-{\rm Y})dy+ ({\rm R}-{\rm Z})dz= d{\rm V}\;,
$$
 to be integrable; in addition, the integral V must
be a function of $q$. Since $\frac{dp}{q}=d{\rm V}$ or $dp=qd{\rm V}$, 
it is clear that the
pressure $p$ cannot have a definite value unless the expression $
 qd{\rm V}$ 
can
be integrated. However, it should also be noted that in this case it
is not necessary that the expression 
$$
({\rm P}-{\rm X})dx+({\rm Q}-{\rm Y})dy+ ({\rm R}-{\rm Z})dz
$$
be integrable, only that on being
multiplied by a certain function U it becomes integrable. Thus, let
$$
{\rm U}({\rm P}-{\rm X})dx+{\rm U}({\rm Q}-{\rm Y})dy+ 
{\rm U}({\rm R}-{\rm Z})dz= d{\rm W}\;,
$$
since $\frac{dp}{q}= \frac{d{\rm W}}{\rm U}$,  or $dp=\frac{qd{\rm
    W}}{\rm U}$ for this equation to be possible it is
sufficient that W be a function of $\frac{q}{\rm U}$, or that W be a function of
zero dimension of the quantities $q$ and U.\footnote{This latter
  expression is equivalent, in 18th century terminology, to the
  condition that W should depend only on the ratio $q/U$.}\\

{\bf 37}.~~~In general, however the elasticity $p$ depends on the density
$q$ or on some other property denoted by $r$ which is any function of
the coordinates $x$, $y$, $z$ that could also contain time $t$, it is clear
from our equation $q=\frac{dp}{\rm dV}$  that the differential $dp$ 
must always be
divisible by $d{\rm V}$, where $d{\rm V}$ denotes not so much a total
differential than the expression 
$$
({\rm P} - {\rm X})dx + ({\rm Q} - {\rm V})dy + ({\rm R} - {\rm
  Z})dz\;,
$$ 
and this so much that, as a result of division the differentials $dx$, $dy$ and
$dz$ are entirely eliminated from the calculations, because both $p$ and $q$
must always be expressed in terms of finite functions of the
variables $x$, $y$ and $z$, without their differentials entering into these
functions. Now this could not be so unless there were a function ${\rm
  U}$,
multiplication by which rendered the expression $d{\rm V}$ integrable:
indeed, setting $\int {\rm U}d{\rm V} = {\rm W}$, clearly, $p$ must
be a function of ${\rm W}$ in order for the expression $\frac{dp}{\rm
  dV}$ to take a definite
value corresponding to the density $q$.\\

{\bf 38}.~~~Since ${\rm U}\,d{\rm V}=d{\rm W}$, we have $q= \frac{{\rm
 U}dp}{d{\rm W}}$. Consequently, if we choose ${\rm W}$ to be any
 function of the coordinates $x$, $y$ and $z$, which contains time $t$
 among the constants, and if we set $p$ equal to any function of
 ${\rm W}$, namely\footnote{For representing the functional
 dependence, now denoted $f(x)$, \label{fn:functions} Euler used the
 notation $f,x$ or $f:x$. For example Euler's $\varphi,{\rm W}$ and
 $\varphi',{\rm W}$ would now be denoted $\varphi({\rm W})$ and
 $\varphi'({\rm W})$. In Euler, 1755c, the comma is omitted.
}
 $p=\varphi,{\rm W}$, and $dp =d{\rm W}.\varphi',{\rm W}$, we will
 have $q={\rm U}.\varphi',{\rm W}$, whence ${\rm U}=
 \frac{q}{\varphi',{\rm W}}$.  Thus, in however way the density $q$ is
 expressed in terms of the elasticity $p$ and some other function $r$
 of the coordinates $x$, $y$ and $z$, we obtain the value ${\rm
 U}=\frac{q}{\varphi',{\rm W}}$ and, consequently, the value $ d{\rm
 V}= \frac{d{\rm W}.\varphi',{\rm W}}{q}$, which then gives us the
 following equation:
$$
({\rm P} - {\rm X})dx + ({\rm Q} - {\rm V})dy + ({\rm R} - {\rm
  Z})dz = \frac{d{\rm W}.\varphi',{\rm W}}{q}= \frac{dp}{q}\;.
$$ 
This will yield the values of X, Y, Z, from which we must 
then look for  the values 
of the velocities $u$, $v$ and $w$:  and when 
the latter also satisfy the continuity condition, we shall have a case of possible motion of the fluid.\\ 

{\bf 39}.~~~The question of the nature of the
expression $({\rm P} - {\rm X}) dx + ({\rm Q} - {\rm Y})dy + ({\rm R}
- {\rm Z})dz$ then reduces to the following. When the
density $q$ is constant or depends only on the elasticity $p$, this
expression must be absolutely integrable and to this end one must
determine suitable values of the three velocities $u$, $v$ and
$w$.When the density $q$ depends on a given function of
place and time,\footnote{The function $r$.} the expression must be such that it becomes
integrable on multiplication by some given function U. In both cases,
then, the velocities $u$, $v$ and $w$ must be such that the equation 
$$
({\rm P}-{\rm X })dx+({\rm Q}-{\rm Y})dy+({\rm R}-{\rm Z})dz =0
$$
be
soluble;\footnote{Indeed, if  $\Phi(x,y,z)= {\rm Cnst.}$ is the
  general integral of this equation, then the form $({\rm P}-{\rm X
  })dx+({\rm Q}-{\rm Y})dy+({\rm R}-{\rm Z})dz$ must vanish whenever
  the differential $d\Phi$ vanishes; hence the two forms are
  proportional, which means that there exists an integrating factor
  for the first form.}  and we  know the conditions under which a differential
equation with three variables is soluble; having satisfied these
conditions, it remains to satisfy that imposed by continuity.

{\bf 40}.~~~These are the conditions which restrict the functions expressing the three
velocities $u$, $v$ and $w$, and the study of the motion
of fluids reduces to determining, in general form, the nature of those
functions such that the conditions of our differential equation and
of continuity be fulfilled. Since the quantities X, Y and Z depend not only on the
velocities $u$, $v$ and $w$ themselves but also on their variability with
respect to each of the coordinates $x$, $y$ and $z$ and, moreover, on time
$t$, this study would appear to be the most far-reaching of those to be
encountered in the field of Analysis, and if we are unable to achieve a
complete understanding of the motion of fluids, it is not Mechanics or the
inadequacy of the known laws of motion but Analysis itself that is to blame,
given that the entire Theory of the motion of fluids has just been reduced to
the solution of analytical equations.\\

{\bf 41}.~~~Since a general solution must be deemed impossible due to the
shortcomings of Analysis, we must content ourselves with the consideration of
certain particular cases, especially as the study of several cases seems to be
the only means of perfecting our knowledge. Now the simplest case imaginable
is, no doubt, that in which the three velocities $u$, $v$ and $w$ are set
equal to zero, namely the case in which the fluid remains at perfect rest and
which I dealt with in my previous Memoir.\footnote{See Euler, 1755a.} The
formulas we have obtained for motion in general also include the case of
equilibrium, since when ${\rm X}=0$, ${\rm Y}=0$ and ${\rm Z}=0$ we have:
$\frac{dp}{q}={\rm P}dx+{\rm Q}dy+{\rm R}dz$, and
$\left(\frac{dq}{dt}\right)=0$, 
 from
which it follows, first of all, that the density $q$ cannot depend on time
$t$,
 i.e.\ should remain always the same at the same place. Furthermore, the forces
${\rm P}$, ${\rm Q}$ and ${\rm R}$ must be such that the differential 
expression ${\rm P}dx + {\rm Q}dy + {\rm R}dz$
either is integrable, when $q$ is constant or depends only on the elasticity $p$,
or becomes integrable upon being multiplied by a suitable function.\\

{\bf 42}.~~~In my Memoir on fluid equilibrium\footnote{Euler, 1755a.} I only considered cases of the
acting forces ${\rm P}$, ${\rm Q}$, ${\rm R}$ for which the differential
expression ${\rm P}dx + {\rm Q}dy + {\rm R}dz$ is integrable, since this
seemed to be the only case that could occur in Nature. In fact, if the density
$q$ is either constant or depends only on the pressure $p$, the fluid could
never be in equilibrium unless this condition relating to the acting forces
is satisfied. However, if it were possible for the acting forces to obey
some other law, there could be equilibrium provided that the forces were such
that there existed some function ${\rm U}$ which when multiplied by the
expression ${\rm P}dx + {\rm Q}dy + {\rm R}dz$ made that expression
integrable, or, equivalently,  provided that the differential equation ${\rm P}dx + {\rm Q}dy +
{\rm R}dz =0$ were integrable; for then if the density $q$ is equated
to this function ${\rm U}$ or to the product of this function
${\rm U}$ and some arbitrary function of the elasticity $p$, equilibrium may
also exist.  However, since these cases may not be possible, I shall not
consider them in greater detail.\\

{\bf 43}.~~~After the case of equilibrium, the simplest state that could exist
in a fluid is that in which the entire fluid is in uniform motion in the same
direction. Let us see, then, how this state is described by our two
formulas. In this case, the three velocities being constant, we set $u =a$, 
$v= b$ 
and $w=c$; we have ${\rm X}=0$, ${\rm Y}=0$ and ${\rm Z} =0$. 
Then our two equations assume the
form:
\begin{eqnarray}
&&\frac{dp}{q} = {\rm P}dx+{\rm Q}dy+{\rm R}dz\;,\nonumber\\
&&\left(\frac{dq}{dt}\right)+a\left(\frac{dq}{dx}\right)+b\left(\frac{dq}{dy}\right)+c\left(\frac{dq}{dz}\right)=0\;,\nonumber
\end{eqnarray}
and hence it is dear that if the density $q$ is constant, the condition of
the second equation is satisfied; however, the first equation cannot be
satisfied unless the expression ${\rm P}dx + {\rm Q}dy + {\rm R}dz$ 
admits integration, just as if
the fluid were at rest.  Of course, such motion can have no effect on the
pressure.\\

{\bf 44}.~~~If, however, the density $q$ is not constant, let us first see what
function of $x$, $y$, $z$ and $t$ it must be for the second equation to be
satisfied. This leads us to the curious analytical question of what
function of the variables $x$, $y$, $z$ and $t$ must be taken for $q$ in order
that: 
$$
\left(\frac{dq}{dt}\right)+a\left(\frac{dq}{dx}\right)+b\left(\frac{dq}{dy}\right)+c\left(\frac{dq}{dz}\right)=0\;.
$$ This would appear to be very difficult to answer if formulated in its
broadest possible form. However, since when $a=0$, $b=0$, $c=0$ the quantity
$q$ is any function of $x$, $y$, $z$ that does not contain time $t$, if we
reduce this case\footnote{The case of motion.} to that of rest by imposing on the volume an equal and
opposite motion, then, clearly, after time $t$ the coordinates $x$, $y$ and
$z$ will be transformed by the change into $x - at$, $y - bt$, $z - ct$. From
this we conclude that our equation will be satisfied if as $q$ we take any
function of the three quantities $x - at$, $y - bt$, $z - ct$.\footnote{Here
  Euler performs a Galilean transformation.} And in fact it
is easy to see that such a function satisfies the equation, since 
$$
dq={\rm L}(dx - adt) + {\rm M}(dy - bdt) + {\rm N}(dz - cdt)\;,
$$ 
and, consequently,
\begin{eqnarray}
\left(\frac{dq}{dt}\right)&=&-a{\rm L} -b{\rm M}-c{\rm N}; \quad
\left(\frac{dq}{dx}\right)={\rm L}\;;\nonumber\\ 
\left(\frac{dq}{dy}\right)&=&{\rm M}\;;\quad\qquad {\rm and} \qquad 
\left(\frac{dq}{dz}\right) ={\rm N}\;.\nonumber
\end{eqnarray}\\

{\bf 45}.~~~Now, as I have already noted, in order to satisfy the first equation
it is necessary that after multiplication by some function U the differential
expression ${\rm P}dx + {\rm Q}dy + {\rm R}dz$ be integrable.  Therefore let
$\int U({\rm P}dx + {\rm Q}dy + {\rm R}dz)= {\rm W}$, where the constant of
integration also in some way contains time $t$. Clearly, the expression ${\rm
P}dx + {\rm Q}dy + {\rm R}dz$ will also be integrable if it is multiplied by
${\rm U}f,{\rm W}$,\footnote{The equivalent modern notation would be ${\rm
U}f({\rm W})$, cf. footnote~\ref{fn:functions}.
}  where U and W are known
functions, since the acting forces are assumed to be known. Thus, if $q$ does
not depend on $p$, then necessarily $q = {\rm U}f,{\rm W}$, whence the
function of the three quantities $x - at$, $y - bt$ and $z - ct$ must be so
determined that it can be reduced to the form ${\rm U}f,{\rm W}$.  If, however,
$q$ depends only on $p$, the expression ${\rm P}dx + {\rm Q}dy + {\rm R}dz$
must be absolutely integrable or ${\rm U}= 1$; then, since $p$ will be found
in the form of a function of W, the density $q$ will likewise be a function of
W, which must also be a function of the quantities $x - at$, $y - bt$ and $z -
ct$, and from this we can deduce the nature of this function.\\

{\bf 46}.~~~However, it can be seen that, in general, the pressure $p$ must
always be a function of W, since otherwise the density could not be a finite
function. Therefore let $p=f,{\rm W}$ and $dp=d{\rm W}.f',{\rm W}$; then, by
virtue of the fact that ${\rm P}dx + {\rm Q}dy + {\rm R}dz = \frac{d{\rm
W}}{\rm U}$, we obtain $q={\rm U}f',{\rm W}$.  Consequently, this case could
not arise unless the density $q$ was proportional to the product of the
quantity U and a function of the pressure $p$ or to the product of the
quantity ${\rm U}\varphi,{\rm W}$ and any function of $p$, where
$\varphi,{\rm W}$ is used to denote a given function of W.  For example, let
$q= pp{\rm U} \varphi,{\rm W}$; we then have $f',{\rm W}= \frac{d(f,{\rm
W})}{d{\rm W}} = (f,{\rm W})^2 \varphi,{\rm W}$,\footnote{The
  equivalent modern form would be  $f'({\rm W})= df({\rm W})/d{\rm W}=
f^2({\rm W})\varphi({\rm W})$.
} whence we find that the
unknown function $f,{\rm W}$ is composed of W, for in this example we have
$\frac{1}{f,{\rm W}} = -\int d{\rm W},\varphi {\rm W} =\frac{1}{p}$ and hence
$p$ can be expressed in terms of W and thus, the quantity $q$ will also be
known. When the latter can be reduced to the form of a function of $x - at$,
$y - bt$ and $z - ct$, the assumed state of the fluid will be possible and we
shall know the pressure and the density at any time and at any point.\\

{\bf 47}.~~~An example\footnote{In this example forces are considered which do
  not derive from a potential and the integrating 
factor U is found for these forces.}  will
 throw more light on these operations which, as they are not yet sufficiently
 familiar, might appear overly obscure. Thus, let ${\rm P} =y$, ${\rm Q} = -x$ 
and ${\rm R} = 0$;
 since $\frac{dp}{q}= ydx-xdy$,  we obtain ${\rm U}= \frac{1}{yy}$ and
${\rm W}= \frac{x}{y} +{\rm T}$, where ${\rm T}$ is any function of time
 $t$. Moreover, let $q=\frac{pp}{yy}$; since $\frac{dp}{pp}= \frac{ydx - x
 dy}{yy}$, we shall obtain $\frac{1}{p} = \Theta -\frac{x}{y}$, and $p=
 \frac{y}{\Theta y -x}$, where the constant $\Theta$  also contains time $t$. 
As a result, we
 have $q= \frac{1}{(\Theta y-x)^2}$, and this expression must be a function 
of $x - at$ and $y - bt$ , since $z$ does
 not enter into it and this is only possible when $\Theta = \frac{a}{b}$; we 
then have $q=\frac{bb}{(ay-bx)^2}$, and $p=\frac{by}{ay-bx}$. Thus,
 neither the pressure nor the density depend on time and at a given point will
 be always the same. This example shows how the calculations should be
 performed in other cases that might be imagined.\\

{\bf 48}.~~~Having dealt with this case in which the three velocities are
constant, let us now assume that two velocities $u$ and $v$ vanish, which
corresponds to the case in which all the fluid particles move in the direction
of the OA axis, so that the trajectory described by each is a straight
line\footnote{This is the case of so-called shear flow.} parallel to the OA
axis;  this case differs from the previous one, since the velocity  $u$ is 
assumed to vary with respect to both place and time. Since 
$$
{\rm X}=\left(\frac{du}{dt}\right)+u\left(\frac{du}{dx}\right);\quad{\rm Y}=0;  
\quad{\rm Z} =0\;,
$$
our two
equations will take the form:

\begin{eqnarray}
\frac{dp}{q}&=& {\rm P}dx+ {\rm Q}dy+{\rm R}dz-dx\left(\frac{du}{dt}\right) -
udx\left(\frac{du}{dx}\right)\;,\nonumber\\
&\& & \quad\left(\frac{dq}{dt}\right)+\left(\frac{d.qu}{dx}\right)=0\;.\nonumber
\end{eqnarray}
This latter equation tells us, first of all, that the expression $qdx - qudt$ 
must be integrable, the quantities $y$ and $z$ being 
considered constant with respect to this integration. Thus, the product of $q$ 
and $dx - udt$ must be a total differential,  i.e.\ must be integrable.\\

{\bf 49}.~~~If the density of the fluid is everywhere and always the same, i.e.\ 
if  $q$ is a constant equal to $g$, then, since
$\left(\frac{du}{dx}\right)=0$, 
it is dear that the velocity
 $u$ must be independent of the variable $x$. Let $u$ be any function of  
the  two coordinates $y$, $z$ and time $t$. Then our differential equation 
will take the  form: 
$$
\frac{dp}{q}= {\rm P}dx+ {\rm Q}dy+{\rm R}dz-dx\left(\frac{du}{dt}\right)\;,
$$
where time $t$ is assumed to be constant; thus, this expression must be
 integrable. Accordingly, if the expression ${\rm P}dx + {\rm Q}dy + 
{\rm R}dz$ obtained from
 considering the acting forces is integrable in itself, then 
$dx\left(\frac{du}{dt}\right)$  must
 also be integrable.  The expression $\left(\frac{du}{dt}\right)$ does not contain $x$, but if it were
 to contain $y$ and $z$, the expression $dx\left(\frac{du}{dt}\right)$ 
could not be  integrable. Thus, $\left(\frac{du}{dt}\right)$
 must not contain $y$ and $z$. Let Z be any function of $y$ and $z$, and T 
any  function of time $t$ only; then the quantity $u ={\rm Z} + {\rm T}$ 
will satisfy this
 condition, whence by virtue of the fact that ${\rm P}dx+ {\rm Q}dy+{\rm R}dz
= d{\rm V}$ and $\left(\frac{du}{dt}\right)=\left(\frac{d{\rm T}}{dt}\right)$,
we obtain the following integral: $\frac{p}{q}= {\rm V} -
x \left(\frac{d{\rm T}}{dt}\right) + {\rm Cnst}$.\\

{\bf 50}.~~~In further clarification of this case, it should be noted that each fluid
particle Z moves exclusively in the direction ZP parallel to the ZA axis and
hence the motion of each fluid element will describe a straight line parallel
to that axis, so that for the same element there is no change in the value of
the two coordinates $y$ and $z$. Thus, the motion of each particle will either be
uniform or will vary with time in such a way that at each instant all the
particles undergo the same changes in their motions, which is obvious from the
equation $u={\rm Z} + {\rm T}$.  As to the state of pressure, given
that we have  $p=g{\rm V} -gx\left(\frac{d{\rm T}}{dt}\right) + {\rm Cnst.}$
where the constant has any  dependence on time $t$, it depends not 
only on the effort\footnote{See footnote~\ref{fn:effort}.} 
${\rm V}$ but also on the change of
velocity undergone by each element of the fluid;  and, moreover, it may vary in
any way with time.\\

{\bf 51}.~~~This case provides me with an opportunity to deal with certain
questions which naturally arise and whose clarification is of the
utmost importance for the theory of both fluid equilibrium and fluid
motion. First of all, surprisingly, a change in the velocity of the
fluid can occur without the acting forces ${\rm P}$, ${\rm Q}$, ${\rm
  R}$ 
helping to produce
it. Since such a change could take place even when the acting forces
vanish, it is reasonable to inquire how it is produced. Next, it also
seems paradoxical that the pressure can vary arbitrarily at any
instant, and that irrespective of the aforesaid change to which the
motion is subjected. The latter difficulty remains even in the state
of equilibrium. Thus, letting the three velocities $ u$, $v$,  
$ w$ vanish, for
incompressible fluids we have the integral $\frac{p}{g} = {\rm V} +
{\rm Cnst.}$, where the constant may  contain the time $t$ in any
way.\\

{\bf 52}.~~~To understand this more clearly, one need only imagine a
certain mass enclosed in a vessel. Clearly, the state of pressure
depends not only on the acting forces but also on any extraneous
forces 
which might be exerted on the vessel. For, even if there were no
acting forces, by means of a piston applied to the fluid one could
successively produce every possible state of pressure without the
equilibrium being disturbed. This is precisely what we can conclude
from our formula, which in this case shows that $\frac{p}{g}$ is a
function of time $t$. From this we see that the state of pressure may vary at any
instant, irrespective of the equilibrium. However, if for each instant 
of time the pressure at any point is known, then the pressures at
all the other points can be determined, and since the force applied
to the piston might now increase and now decrease, the calculations
must reflect all these possible changes. The same variability should
also be observed when the fluid is subjected to the action of arbitrary
accelerative forces, so that at each instant the state of pressure is
indeterminate and depends on the force then acting on the piston.\\

{\bf 53}.~~~Here, then, is a vital difference between the accelerative
 forces, which act on all the elements of the fluid, and the force of
 a piston that presses on the fluid. Only the accelerative forces
 enter 
into our differential equation, while the piston force enters
 into the calculations only after integration and only affects the
 constant of integration.  Consequently, in each case the constant
 must be so determined that at the point at which the piston acts the
 pressure is exactly equal to the force driving the piston at each
 instant, and it is for this reason that the constant contains time,
 so that it can be varied with time at will, as the circumstances
 require. This variability can always be represented by the action of
 a piston since, whatever the nature of the case considered, for it
 to be determined it must always be assumed that at one point at least
 in the fluid the pressure is known at every instant, and it is
 precisely this which makes it possible to determine the constant
 introduced into the calculations through  the integration of
 our differential equation.\\

{\bf 54}.~~~However, in our case of the motion considered in \S~49, let
 us also assume that the accelerative forces vanish, i.e.\ that ${\rm V}=0$,
 and to make this case perfectly determinate, let us assume that $u =a
 + \alpha y + \beta t$.\footnote{In Euler, 1755b the symbol $\beta$ is used in
 the r.h.s. of this equation; in the printed version it is replaced by a
 symbol resembling a capital C with curled ends.} Then the equation for the pressure will take the form
 $\frac{p}{g} ={\rm Cnst.} - \beta  x$. Let us assume, moreover, that this constant is
 equal to $\gamma +\delta t$, so that  $\frac{p}{g} = \gamma +\delta t
 -\beta  x$, and let us see under what
 conditions this motion can take place. Since each fluid element
 moves in the direction of the OA axis, the motion could only take
 place in a cylindrical pipe laid in the same direction. Let ABIO
 (Fig.~4) be that pipe and initially, at $t=0$, let the fluid occupy
\begin{figure}[!h]
  \includegraphics[scale=0.5,angle=-2]{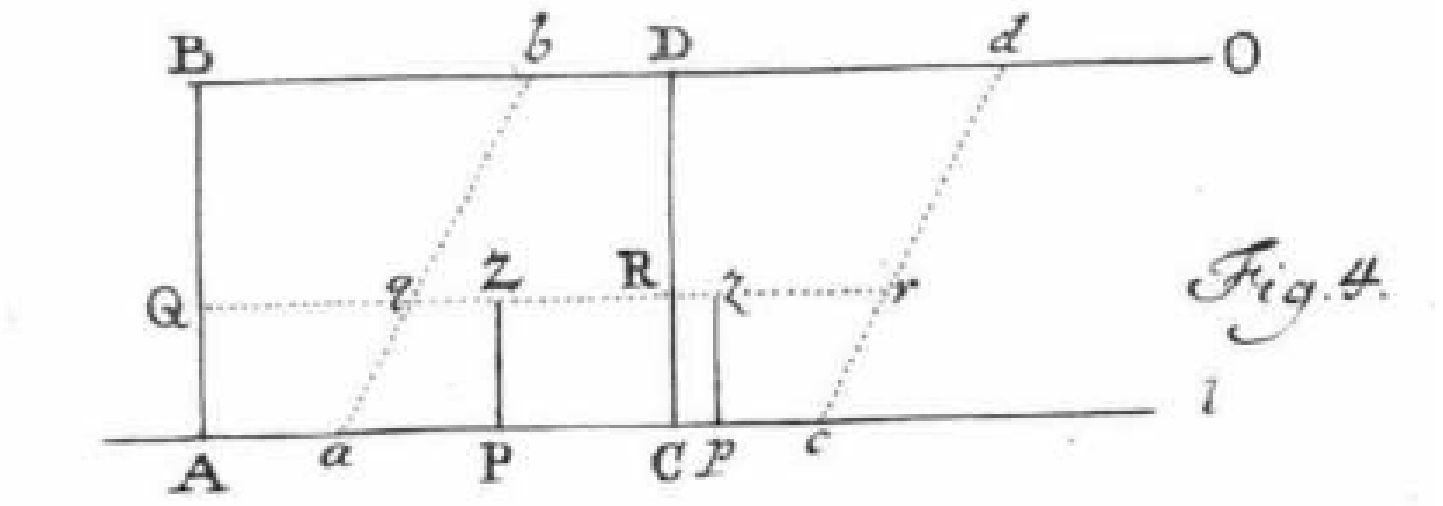}%
\end{figure}
the portion ABCD bounded by cross sections AB and CD perpendicular
 to the pipe. We will reckon the abscissas from the point A along the
 straight line AI and let the pressure $p$ be equal to $\gamma g$  
everywhere along
 the base AB and to $\gamma g -\beta  g.{\rm AC}$  along the other
 base CD. 
In the
 interior of the fluid, however, at any point Z with the coordinates
 ${\rm AP}=x$, ${\rm PZ}=y$ , the pressure will be equal to $\gamma
 g-\beta gx$. Consequently, it
 is impossible to consider the fluid in the pipe beyond CD, 
 taking ${\rm AC} = \frac{\gamma}\beta $, so that the pressure at CD does not
 become negative.\\

{\bf 55}.~~~Let us set  for this determinate fluid mass the length
 ${\rm AC}= b$ and the width ${\rm AB} = {\rm CD} =c$, the height not entering into
 consideration since neither the velocities nor the pressures depend
 on the third coordinate $z$; when $\gamma = \beta  b$, in the
 initial 
state ABCD
 the pressure is equal to $\beta bg$  on the base AB and zero on the base CD,
 while at any point Z it is equal to $\beta g(b-x) =\beta g.{\rm
 CP}$. 
We will assume
 that in this state the fluid has a motion in the direction of the
 pipe such that the velocity on the line AC is equal to $a$ and that on
 the line BD equal to $a+\alpha c$, while on any line QR parallel to the
 direction of the pipe it is equal to $a + \alpha y$, where ${\rm
 AQ}={\rm CR}=y$. Thus , we believe
 that something has caused this motion to be impressed on the
 fluid and that, at the initial instant, the surface AB is subjected
 to the said force $\beta bg$, exerted by means of a piston, 
while the other base
 CD is not subjected to any pressure. However, at subsequent moments
 of time the forces acting on the end faces could vary
 arbitrarily. Now this variability is determined by the hypotheses we
 have just established.  Therefore let us see how by virtue of these
 hypotheses the motion of the fluid will be continued.\\

{\bf 56}.~~~After the lapse of a time $t$, all the fluid elements on the
 line QR will have a velocity in that same direction equal to $a +
 \alpha y +\beta t$, as a result of which in the time $dt$ they will
 travel a distance $(a + \alpha y +\beta t)dt$; thus, from the
 beginning of the motion they will have traveled a distance $at +
 \alpha yt + {\scriptstyle \frac{1}{2}}\beta tt$; and the alignment
 of fluid particles\footnote{Euler uses ``fil\'ee du fluide'' where
 ``fil\'ee'' is a somewhat poetic variant of ``file'' (alignment,
 file) or ``fil'' (thread); this is just a line of fluid elements and not what is now called
 a fillet of fluid, the latter having also an infinitesimal width, a
 concept introduced by Euler, 1745 (see Grimberg, Pauls and Frisch,
 2008).
}  initially at QR will now have advanced to $qr$, having
 traversed the distance ${\rm Q}q=at + \alpha yt + {\scriptstyle
 \frac{1}{2}}\beta tt$. Thus, the thread AC will have arrived at
 $ac$, having traveled a distance ${\rm A} a=at + {\scriptstyle
 \frac{1}{2}}\beta tt$, while the thread BD will have arrived at
 $bd$, having traveled a distance ${\rm B}b = at + \alpha ct +
 {\scriptstyle \frac{1}{2}}\beta tt$, so that the fluid mass will
 now be bounded by the faces $ab$ and $cd$, which are straight but
 inclined to the direction of the pipe. The pressure on the face $ab$ at
 $q$ must now be $g(\beta  b +\delta t -\beta .{\rm Q}q) = g({\beta} b +\delta t -\beta  at -\alpha \beta  yt - {\scriptstyle
 \frac{1}{2}}
\beta \beta tt)$, and on the face $cd$ at $r$ it must now be
$g(\beta  b +\delta t -\beta .{\rm Q}r) = g(\delta t -\beta  at
 -\alpha \beta  yt - {\scriptstyle  \frac{1}{2}} \beta \beta tt)$.
Thus, we need to visualize pistons which
 act with these forces on the two end faces $ab$ and $cd$, and since the
 pressures are not the same over the entire length of these faces, the
 pistons must be imagined as being flexible and pliable enough to
 exert such pressures.\\

{\bf 57}.~~~This motion would remain the same if in integrating the pressure $p$
we were to take any function of $t$ instead of $\delta t$, but then the state of
pressure in the fluid mass would be different at each instant of time, even
though the assumed motion of the fluid itself would not be affected in any
way. Thus, let us set $\delta t = \beta  at + \alpha \beta ct +
{\scriptstyle \frac{1}{2}} \beta \beta tt$; after a time $t$ the pressure at any
point $q$ on the face $ab$ will be $g[\beta  b + \alpha \beta (c-y)t]$,
and at any point $z$ on the line $qr$ it will be equal to $g[\beta  b + \alpha
\beta (c-y)t -\beta .qz]$; therefore the pressure at the other end $r$
will be $ \alpha \beta g(c-y)t$. Hence, on the face $ab$ the pressure will
be equal to $\beta g(b + \alpha ct)$ at $a$ and to $\beta gb$ at $b$,
while on the other face $cd$ the pressure will be equal to ${\alpha \beta}gct$ at $c$ and to zero at $d$. Moreover, each thread QR will move in its
own direction with uniform acceleration, i.e.\ will receive equal increments
of velocity in equal times. The study of this particular case could serve to
elucidate the calculations to be made in all other cases.\\

{\bf 58}.~~~Let us now return to the case proposed (\S~48) and assume the
density $q$ to be constant and equal to $g$, while making the forces ${\rm P}$,
${\rm Q}$, ${\rm R}$ such that the fluid could never be in equilibrium. To
this end, let ${\rm P}=0$, ${\rm Q} = -\frac{x}{a}$ and ${\rm R}
=-\frac{x}{a}$ and let $u= b+\frac{(y+z)t}{a}$, so that we have
$\left(\frac{du}{dx}\right) =0$ and $\frac{dp}{g} = -\frac{xdy +xdz}{a}
-\frac{ydx+zdx}{a}$,\footnote{In the printed version the two fractions in the
  r.h.s. have a minus instead of a correct plus in the numerator; in the 
manuscript  Euler, 1755c, the handwritten notation is ambiguous.} whence by integration we obtain $\frac{p}{g} = {\rm
Cnst.}-\frac{xy+xz}{a}$, where the constant may contain time in any way. Thus,
it is not possible for the entire fluid mass ever to remain at rest, since
even if we set $b=0$ in order to have the fluid at rest at the outset when
$t=0$, immediately after that first instant it would be agitated and only the
elements for which $y=0$ or $z=0$ or $y + z=0$ would remain at rest; all the
others would be set in motion either forward or backward, depending on whether
$y + z$ was positive or negative. It is also easy to determine the pressures
required to maintain the assumed motion.\\

{\bf 59}.~~~Let, however, the density be no longer constant but variable, i.e.\ let
the fluid be compressible. Then in order for the expression $qdx - qudt$ to be
a total differential we can take for $u$ any function of the variables $x$, $y$, $z$
and $t$.  Here, since only $x$ and $t$ are regarded as variable, while
$y$ and $z$ are taken constant, it will always be possible to assign a
quantity $s$ such that $s(dx - udt)$ is integrable. Let S be that integral; 
then
this condition will be satisfied if we take $q=sf:{\rm S}$,\footnote{This
  equation would now be written $q= sf({\rm S})$.
} Furthermore, it is now necessary that the following differential 
be integrable:
$$
\frac{dp}{q} = {\rm P}dx+{\rm Q}dy+{\rm R}dz-dx \left(\frac{du}{dt}\right)
-udx\left(\frac{du}{dx}\right)\;.
$$
Note that if the forces P, Q, R were to vanish, the
pressure $p$ would become a function of $x$ and $t$ and hence the quantity 
$q\left(\left(\frac{du}{dt}\right) +u\left(\frac{du}{dx}\right)\right)$
would only involve the two variables $x$ and $t$, from which the nature of the
function $u$ must be determined, insofar as it involves $y$ and $z$.\\

{\bf 60}.~~~Although I have assumed that $v=0$ and $w=0$, these formulas cover
all the cases in which all the fluid particles always move in the same
direction, the only requirement being that the OA axis be taken in
that direction. Therefore we will also be able to solve our equations
when the direction of motion is inclined to the three axes, which
cannot fail to throw further light on the analysis. To this end, let
us consider the true velocity of any fluid particle Z and let that
velocity be equal to $\ttaurus $, and since its direction is given with
respect to the three axes, the velocity components will hold certain
ratios to it. Let $u = \alpha {\taurus \kern-1.8ex\taurus}$, 
$v=\beta  \ttaurus$ and 
$w= \gamma \ttaurus$; setting 
$d\ttaurus={\rm K}dt+{\rm L}dx+{\rm M}dy+{\rm
  N}dz$, we shall have
\begin{eqnarray}
{\rm X} &=&\alpha{\rm K}+ \alpha \alpha{\rm L}+\alpha\beta {\rm
  M}+\alpha \gamma{\rm N}\nonumber\\
{\rm Y} &=&\beta {\rm K}+\alpha \beta {\rm L}+{\beta \beta}{\rm M}+ \,\beta \gamma{\rm N}\nonumber\\
{\rm Z} &=&\gamma {\rm K}+\alpha\gamma{\rm L}+\beta \gamma {\rm
  M}+\,\gamma \gamma{\rm N}\;.\nonumber
\end{eqnarray}
Consequently, if, for conciseness, we write
${\rm K} +\alpha{\rm L}+\beta {\rm M} +\gamma {\rm N}={\rm O}$, having
${\rm X}=a{\rm O}$, ${\rm Y}= \beta {\rm O}$, ${\rm Z}= \gamma{\rm O}$, our equations will
take the form:
\begin{eqnarray}
\frac{dp}{q}=&& \!\!\!\!{\rm P}dx\,+\,{\rm Q}dy\,+\,{\rm R}dz\,- \,{\rm O}(\alpha dx
\,+\,\beta dy \,+\,\gamma dz) \nonumber\\
&&\!\!\!\!\!\!\!\!\left(\frac{dq}{dt}\right)+\alpha\left(\frac{d.q\ttaurus
  }{dx}\right)+\beta  \left(\frac{d.q\ttaurus}{dy}\right)
+\gamma\left(\frac{d.q\ttaurus}{dz}\right)=0\;.\nonumber
\end{eqnarray}\\

{\bf 61}.~~~First, let the density $q=g$. As we have seen in \S~44, in
order to satisfy the equation $\alpha\left(\frac{d\ttaurus
   }{dx}\right)+\beta  \left(\frac{d\ttaurus
   }{dy}\right)+\gamma \left(\frac{d\ttaurus
   }{dz}\right)=0$ the quantity $\ttaurus$ must be any function of
the quantities $\alpha y - \beta x$  and $\alpha z - \gamma x$ or 
$\beta z -\gamma y$ and, in addition, may in an arbitrary
 way contain time $t$. Thus, let $\ttaurus $  be any function of the
quantities $\alpha y - \beta x$, $\alpha z - \gamma x$, and $t$,
since the expression $\beta z -\gamma y$ has
already been formed from the other two. From this it is easy to see
that at each instant the velocity of particles  on the same straight
line parallel to the direction of motion will be everywhere the same,
just as the nature of the hypothesis requires. Hence the differential
of $\ttaurus $ will have the following form:
$$
d\ttaurus= {\rm F}dt + {\rm G}(\alpha dy -\beta 
dx) +{\rm H}(\alpha dz -\gamma dx)\;,
$$ 
so that ${\rm K}={\rm F}$; ${\rm L}= -\beta {\rm G}-\gamma {\rm H}$;
${\rm M}=\alpha {\rm G}$; and ${\rm N}= \alpha {\rm H}$. Consequently,
${\rm O}={\rm F}$ is a function of $\alpha y-\beta x$, $\alpha z
-\gamma x$ and of $t$. Hence the differential equation, 
which remains to be solved, will be :
$$
\frac{dp}{q} = {\rm P}dx={\rm Q}dy+{\rm R}dz -{\rm F}(\alpha dx
+\beta  dy + \gamma dz)\;.
$$\\

{\bf 62}.~~~The time $t$ being here assumed constant, if the expression
${\rm P}dx + {\rm Q}dy + {\rm R}dz=d{\rm V}$ is integrable in itself,
the other part of the equation ${\rm F}(\alpha dx +\beta  dy +
\gamma dz)$ must be likewise, and this could not be so unless F
were a function of $\alpha x +\beta  y + \gamma z$ and of time $t$.
In addition, however, F must must also be a function of the quantities
$\alpha y-\beta x$, $\alpha z -\gamma x$ and time $t$;
consequently, since the expression $\alpha x +\beta  y + \gamma z$
cannot be formed from the expressions $\alpha y-\beta x$ and $
\alpha z -\gamma x$, it is clear that the quantity F must be a
function of time $t$ only.  Consequently, the velocity $\ttaurus
$ will have the form $\ttaurus =
{\rm Z}+{\rm T}$, where Z denotes an arbitrary function of the two
quantities $\alpha y-\beta x$ and $\alpha z -\gamma x$ that does not
contain time $t$, while T is an arbitrary function of time $t$ only,
so that $d{\rm T}={\rm F}dt$. Hence the integral of our differential
equation will be $\frac{p}{g} = {\rm V}-{\rm F}(\alpha x +\beta  y
+\gamma z) + {\rm Cnst.}$, where the constant may contain time $t$ in
an arbitrary way. Together with the relation $\ttaurus
= {\rm Z} + {\rm T}$, this integral contains
everything relating to the motion in the case in question.\\

{\bf 63}.~~~But if the density $q$ is not constant, it will be
important to obtain the solution of the following equation: 
$$
\left(\frac{dq}{dt}\right)+ \alpha\left(\frac{d.q\ttaurus
    }{dx}\right)+ \beta \left(\frac{d.q\ttaurus
   }{dy}\right)+ \gamma
\left(\frac{d.q\ttaurus}{dz}\right)=0\;.
$$
 However
difficult this may appear, reduction to the previous case shows that 
the velocity $\ttaurus$  can be an arbitrary
 function of the four variables $x$, $y$, $z$ and $t$, while the value of $q$ must be determined as follows. Let us
consider, generally,  an expression 
$$
s(l dx+mdy+ ndz - \ttaurus dt) = d{\rm S}\;,
$$
which has become
integrable after multiplication by $s$, and let $q=sf:{\rm S}$; then, if we set
$d.f:{\rm S} = d{\rm S}.f':{\rm S}$,\footnote{Here, $q=sf:{\rm S}$ and
  $d.f:{\rm S} = d{\rm S}.f':{\rm S}$ would now be denoted $q=sf({\rm
    S})$ and $df({\rm S}) = d{\rm S}f'({\rm   S})$, respectively.
}
our expression will take the form
\begin{eqnarray}
f\!\!\!\!&:&\!\!\!\!{\rm S}\left(\frac{ds}{dt}\right)-sf':{\rm S}.s\ttaurus
 \nonumber\\
&+& \alpha sf:{\rm S}\left(\frac{d\ttaurus }{dx}\right)
+\alpha \ttaurus f:{\rm
  S}\left(\frac{ds}{dx}\right)
+\alpha \ttaurus sf':{\rm S}.ls\nonumber\\	
&+& \beta sf:{\rm S}\left(\frac{d\ttaurus  }{dy}\right)
+\beta  \ttaurus  f:{\rm
  S}\left(\frac{ds}{dy}\right)
+\beta  \ttaurus sf':{\rm S}.ms\nonumber\\	
&+& \gamma sf:{\rm S}\left(\frac{d\ttaurus }{dz}\right)
+\gamma \ttaurus f:{\rm
  S}\left(\frac{ds}{dz}\right)
+\gamma \ttaurus sf':{\rm S}.ns\nonumber
\end{eqnarray}
which must be equal to zero.\\

{\bf 64}.~~~First of all, we equate to zero the terms containing $f':{\rm
  S}$, as a
result of which we obtain $1= \alpha l +\beta  m+ \gamma n$; 
after division by $f':{\rm S}$  the
remaining terms give
$$
\left(\frac{ds}{dt}\right)+ \alpha\left(\frac{d.s\ttaurus
   }{dx}\right)+ \beta \left(\frac{d.s\ttaurus
   }{dy}\right)+ \gamma
\left(\frac{d.s\ttaurus }{dz}\right)=0\;,
$$
which is indeed similar to the expression proposed;
however, it should be noted that the integrability of the quantity
$d{\rm S}$ is conditioned by:
\begin{eqnarray}
\left(\frac{d.s\ttaurus
   }{dx}\right)&=&-\left(\frac{d.ls}{dt}\right);
    \quad \left(\frac{d.s\ttaurus
    }{dy}\right)=-\left(\frac{d.ms}{dt}\right);\nonumber\\
\left(\frac{d.s\ttaurus}{dz}\right)
&=&-\left(\frac{d.ns}{dt}\right)\;;\nonumber
\end{eqnarray}
whence we obtain: $\left(\frac{ds}{dt}\right) (1-\alpha l-\beta m
-\gamma n)=0$,\footnote{The r.h.s. $=0$ is missing both in the
printed version and in Euler, 1755c.}  which is
consistent with the previous condition. Thus, provided that
$\alpha l +\beta  m+ \gamma n=1$, and $s$ is a function such that
$s(l dx + mdy + ndz -\ttaurus dt)=d{\rm S}$, or integrable,
our equation will be satisfied if we
take $q=sf:{\rm S}$, or $\frac{q}{s}$ equal to any function
of S. The
quantities $l$, $m$ and $n$ do not have to be constant, but then the
following
must hold
$$
\alpha \left(\frac{dl}{dt}\right)+ \beta  \left(\frac{dm}{dt}\right)
+\gamma\left(\frac{dn}{dt}\right) =0\;,
$$
a condition already contained in the
equation  $1= \alpha l +\beta  m+ \gamma n$.\\

{\bf 65}.~~~In addition, $l$, $m$ and $n$ must be functions such that the
differential equation $l dx+mdy+ ndz - \ttaurus
dt=0$ becomes possible, since without this
condition it would be impossible to find a multiplier $s$ which made
the equation integrable.  Thus, if we arbitrarily choose some value
for $l$, the values of $m$ and $n$ will be already determined and we
can avoid having to find them. We will set $\alpha l =1$ or $l=
\frac{1}{\alpha}$ ; then, necessarily, $\beta m+\gamma n =0$ and it
remains only to find the factor $s$ for which the expression
$s\left(\frac{dx}{\alpha} -\ttaurus dt\right)$ is
integrable, the two quantities $y$ and $z$ being regarded as
constants. Thus, let ${\rm S}= \int s\left(\frac{dx}{\alpha}
-\ttaurus dt\right)$, so that $y$ and $z$ are
contained in S as constants; we can now take $q=sf\!:\!{\rm S}$, which gives us
the same solution as if we had changed the position of the three
axes so much that one of them coincided with the direction of motion of all
the fluid elements.  Hence we see that this apparent restriction in no
way diminishes the generality of the solution.\\

{\bf 66}.~~~In the same way it would be possible to study several other
 particular cases of sometimes greater and sometimes lesser scope, but
 we would not find a case more general than that in which the three
 velocities $u$, $v$ and $w$ are such that the expression
 $udx+vdy+wdz$ becomes integrable.\footnote{In \S\S~30--33 above, Euler
 has already pointed out the possibility and given examples of
 non-potential fluid flows. Truesdell, 1954 considers that Euler based \S~66
 of his memoir on his previous work (Euler, 1756--1757) which was 
 completed before he had discovered the existence of non-potential
 flows. This seems all the more likely in that, as Truesdell points
 out, Euler here denotes the velocity potential not by ${\rm W}$, as in \S~26,
 but by ${\rm S}$, as in his earlier study.} Let S be an integral which also
 contains time $t$ and let its total differential be  $d{\rm S}=
 udx+vdy+wdz+\Pi dt$.  Since we have
\begin{eqnarray}
\left(\frac{du}{dt}\right)&=&\left(\frac{d\Pi}{dx}\right);\,\, \left(\frac{dv}{dt}\right)=
\left(\frac{d\Pi}{dy}\right);\,\, \left(\frac{dw}{dt}\right)= \left(\frac{d\Pi}{dz}\right);\nonumber\\
\left(\frac{du}{dy}\right)&=&\left(\frac{dv}{dx}\right);\,\, \left(\frac{du}{dz}\right)=
\left(\frac{dw}{dx}\right);\,\, \left(\frac{dv}{dz}\right)=
\left(\frac{dw}{dy}\right),\nonumber
\end{eqnarray}
we shall have
$$
\begin{array}{lcr}
\ds {\rm X}&\ds =&\ds \left(\frac{d\Pi}{dx}\right)+u\left(\frac{d\,u}{dx}\right)+v 
\left(\frac{d\,v}{dx}\right)+w\left(\frac{d\,w}{dx}\right)\nonumber\\[1.9ex]
\ds {\rm Y}&\ds =&\ds \left(\frac{d\Pi}{dy}\right)+u\left(\frac{d\,u}{dy}\right)+v 
\left(\frac{d\,v}{dy}\right)+w\left(\frac{d\,w}{dy}\right)\nonumber\\[1.9ex]
\ds {\rm
  Z}&\ds =&\ds \left(\frac{d\Pi}{dz}\right)+u\left(\frac{d\,u}{dz}\right)+v 
\left(\frac{d\,v}{dz}\right)+w\left(\frac{d\,w}{dz}\right)\nonumber
\end{array}
$$
and our
 differential equation now becomes:
$$
\frac{dp}{q}= {\rm P}dx+{\rm Q}dy+{\rm R}dz -d\Pi -udu-vdv-wdw
$$
 (the last member of which is
 absolutely integrable), while the other equation remains as before:
$$
\left(\frac{dq}{dt}\right) +\left(\frac{d.qu}{dx}\right)+\left(\frac{d.qv}{dy}\right)+\left(\frac{d.qw}{dz}\right)=0\;.
$$\\

{\bf 67}.~~~Thus, everything reduces to finding suitable values for the
 three velocities $u$, $v$ and $w$ that satisfy our two equations, which
 contain everything we know about the motion of fluids. For if these
 three velocities are known, we can determine the trajectory described
 by each element of the fluid in its motion.  Let us consider a
 particle which at a given instant is located at the point Z; for
 finding the trajectory which it has already described and which it
 has yet to describe, since its three velocities $u$, $v$ and $w$ are
 assumed to be known, for its position at the next instant we have $dx
 =udt$, $dy =vdt$ and $dz=wdt$. Eliminating time $t$ from these three equations,
 we obtain two more equations in the three coordinates $x$, $y$ and $z$
 which will determine the unknown trajectory of the fluid element now
 at Z and, in general, we shall know the path which each particle has
 traveled and has yet to travel.\\

{\bf 68}.~~~The determination of these trajectories is of the utmost
importance and should be used to apply the Theory to each case
considered. If the shape of the vessel in which the fluid moves is
given, the fluid particles which touch the surface of the vessel must
necessarily follow its direction; therefore the velocities $u$, $v$
and $w$ must be such that the trajectories derived therefrom lie on
that same surface.\footnote{Here, Euler is drawing attention to the
fact that in order to calculate the motion of a fluid, in addition to
the equations of motion, continuity and state and the initial
conditions, we also need the boundary conditions, namely the vanishing
of the normal component of the velocity.} This makes it quite
clear how far removed we are from a complete understanding of the
motion of fluids and that my exposition is no more than a mere
beginning. Nevertheless, everything that the Theory of Fluids contains
is embodied in the two equations formulated above (\S~34), so that it is
not the laws of Mechanics that we lack in order to pursue this
research but only the Analysis, which has not yet been sufficiently
developed for this purpose. It is therefore clearly apparent what 
discoveries we still need to make in this branch of Science before we
can arrive at a more perfect Theory of the motion of fluids.\\

\end{document}